\documentclass[11pt,a4paper]{article}
\usepackage{jheppub}
\usepackage[paperwidth=22cm]{geometry}
\usepackage[T1]{fontenc}
\usepackage[latin1]{inputenc}
\usepackage{graphicx}
\usepackage{amsmath}
\usepackage{amssymb}

\newcommand{\mf}{\mathfrak}
\newcommand{\ovl}{\overline}
\newcommand{\bl}{\boldsymbol}

\newcommand{\lef}{\left(\,}
\newcommand{\rig}{\,\right)}
\newcommand{\ph}{\phantom}

\newcommand{\eq}{\,=\,}

\begin{document}

\abstract{
In this article we construct and discuss several aspects of the two-component spinorial formalism for six-dimensional spacetimes, in which chiral spinors are represented by objects with two quaternionic components and the spin group is identified with $SL(2;\mathbb{H})$, which is a double covering for the Lorentz group in six dimensions. We present the fundamental representations of this group and show how vectors, bivectors, and 3-vectors are represented in such spinorial formalism. We also complexify the spacetime, so that other signatures can be tackled. We argue that, in general, objects built from the tensor products of the fundamental representations of $SL(2;\mathbb{H})$ do not carry a representation of the group, due to the non-commutativity of the quaternions. The Lie algebra of the spin group is obtained and its connection with the Lie algebra of $SO(5,1)$ is presented, providing a physical interpretation for the elements of $SL(2;\mathbb{H})$. Finally, we present a bridge between this quaternionic spinorial formalism for six-dimensional spacetimes and the four-component spinorial formalism over the complex field that comes from the fact that the spin group in six-dimensional Euclidean spaces is given by $SU(4)$.
}


\title{Two-Component Spinorial Formalism using Quaternions for Six-dimensional Spacetimes}

\author{Jo\'{a}s Ven\^{a}ncio and Carlos Batista}

\affiliation{Departamento de F\'{\i}sica,\\Universidade Federal de Pernambuco,\\50670-901 Recife-PE, Brazil}
\emailAdd{joas.venancio@ufpe.br; carlosbatistas@df.ufpe.br}


\maketitle

\section{Introduction}

The study of spinors is of great physical relevance, inasmuch as fermionic fields are represented by spinorial fields or by tensor products of spinorial fields. Therefore, spinors play a central role in the theory of particles and fields. Particularly, in supersymmetric theories the parameters that label the supersymmetry transformations are always given by spinors. In addition, spinors are of great geometrical relevance, since they carry the fundamental representation of the Lorentz group (or the orthogonal group in the case of a general signature) and, therefore, can be used to build all other representations of this group. In this sense, spinors are the most fundamental objects of a space endowed with a metric.

It is well known that the Lorentz group in four dimensions, $SO(3,1)$, has the special linear group $SL(2; \mathbb{C})$ as its double cover, with $\mathbb{C}$ being the field of the  complex numbers. Such an isomorphism is the basis for the spinorial index notation in four dimensions, which has been introduced in general relativity mainly by R. Penrose and that led to great advances on the comprehension of several physical and geometrical matters \cite{Penrose60, Penrose65, Penrose84}. For example, many results about massless fields propagating in curved four-dimensional spacetimes have been achieved due to the use of the index spinorial formalism \cite{Penrose65}. Likewise, the study of massive particles can also benefit from the use of spinors \cite{Luca14}. As another application for spinorial techniques, scattering amplitudes for Yang-Mills fields have been computed by means of such a tool \cite{Britto:2005fq}.

It turns out that an analogous relation holds in six dimensions, namely the Lorentz group $SO(5, 1)$ has as its double cover also a two-dimensional special linear group, but now over the quaternions \cite{Lounesto,Kugo,Baez:2010bbj}. Thus, we say that $SL(2; \mathbb{H})$ is the spin group in six-dimensional spaces endowed with metrics of Lorentzian signature. In particular, this means that (chiral) spinors are two-component objects, just as happens in four dimensions. The main goal of the present paper is then to establish the basics of this spinorial formalism over the quaternions. Some steps toward this goal have already been taken in Refs. \cite{Kugo,Baez:2010bbj}. Here we make advances like discussing the Lie algebra of the spin group and showing explicitly how a pair of spinors can used to build a light-like vector. Moreover, we discuss how the non-commutative feature of quaternions greatly constrains which tensorial objects carry representations of the spin group $SL(2; \mathbb{H})$. In this work we try to follow a bottom up and self-contained approach, introducing all necessary tools and constructing the intuition step by step. It is also worth pointing out that in ten-dimensional spacetimes chiral spinors can be represented by two component objects, but with entries on octonions \cite{Kugo,Baez:2010bbj}. For some recent applications of this fact to string theory see \cite{Schreiber1,Schreiber2,Schreiber3} and references therein.

As a matter of fact, another spinorial index approach has already been introduced in six dimensions, according to which chiral spinors are represented by four-component objects. This possibility comes from the fact that the double cover of $SO(6)$ is $SU(4)$. Then, by means of the complexification of the Euclidean space, one can represent spinors in Lorentzian signature by four-component objects, as done in Refs. \cite{CaBru,Carlos,Weinberg-Conf6D,Mason6D,Batista_Conf6D,Kerr6D,Bat-Book,Koller83,Howe83}. In the present work we provide the explicit connection between the two-component spinors and the four-component ones.

Quaternions have a long history of applications in physics \cite{Gursey:1950zz}, starting with their utility for handling rotations in three dimensions, since $SU(2)$ is the spin group in three dimensions and its Lie algebra is naturally associated to a representation of the quaternions. Several applications on four-dimensional spacetimes have also been put forward. For instance, in \cite{Morita83} it was shown the quaternioninc formulation of the Dirac theory, in \cite{Morita84} a quaternionic structure of simple supergravity was discussed, while Refs. \cite{Adler86, Adler95} present a quaternionic version of quantum mechanics and quantum field theory. Dirac himself even used quaternions to represent vectors of Minkowski spacetime and used a great deal of ingenuity to write Lorentz transformations as a generalization of a fractional linear transformation over the quaternions \cite{Dirac}. Concerning dimension six, twistors and supertwistors have been investigated using quaternions in Refs. \cite{Bengtsson88,Lukierski91}, while more general aspects of the spinorial formalism using quaternions can be found in \cite{Kugo,Sudbery84}.
For some general remarks on division algebras, see \cite{Cederwall93,Baez:2010bbj}.

There are several reasons for studying spacetimes with dimension greater than four. Indeed, there exists a wide range of physical theories that model our world by means of higher-dimensional spaces \cite{Csaki:2004ay}. The most popular one is string theory, which is a quantum theory that requires our spacetime to have ten dimensions and has the great feature of having spin two excitations in its field content, i.e. quantum gravity is a byproduct of such a theory   \cite{Mukhi:2011zz}. In the string theory context it is often argued that the six extra dimensions, that are not noticed in our daily lives, form a tiny compact submanifold, so that only high energy experiments could probe their existence. It is even possible that higher-dimensional black holes could be generated and detected in particle colliders \cite{Cavaglia}. However, there is also the possibility that non-compact extra dimensions are compatible with the current experimental observations, as long as the fields of the standard model are confined to a four-dimensional brane \cite{ADD1,ADD2,Randall}, with gravity permeating through all dimensions. Higher dimensions are also studied with the intention of learning more about the gravitational field, which can be much more richer when the dimension is greater than four \cite{Emparan}. The dimension of the spacetime has even been considered as a perturbation parameter in the study of black hole dynamics \cite{Emparan2}. More recently, the most important applications of higher-dimensional spacetimes arose in the AdS/CFT context, in which a gravitational theory in $n+1$ dimensions is connected to a field theory living in the $n$-dimensional boundary of the spacetime \cite{Maldacena:1997re,Hubeny:2014bla,Erlich:2005qh}. In particular, this correspondence has led to theoretical predictions on quark-gluon plasma that have been experimentally verified \cite{Finazzo:2014cna,CasalderreySolana:2011us}.
Concerning dimension six, besides being the dimension of the wrapped part of the spacetime in string theory, another important reason to be interested in this dimension is that the conformal group in four dimensions is the orthogonal group of a six-dimensional space, so the study of spinors in six-dimensional spaces are of relevance for the study of conformal field theory in four dimensions \cite{Giardino15}.

 The outline of the present paper is the following. In Sec. \ref{Sec.Quat} we review the basics of quaternionic matrices and introduce a generalization of determinant that is suitable for matrices with quaternionic entries, this is the determinant used to define the group $SL(2;\mathbb{H})$. Then, in Sec. \ref{Sec.Vectors} it is shown the connection between the Lorentz group $SO(5,1)$ and the group $SL(2;\mathbb{H})$. In particular, it is argued that vectors of a six-dimensional spacetime should be represented by $2\times 2$ hermitian matrices over the quaternions. In Sec. \ref{Sec.SpinorRep} we present the fundamental representations of $SL(2;\mathbb{H})$, which define the spinors. These spinorial representations are then used in Sec. \ref{Sec.Tensors} in order to build other representations of the spin group, we also argue that most of the tensorial objects do not carry a representation, due to the noncommutativity of quaternions. In Sec. \ref{Sec.CompVec}, we show how complex vectors are represented in the quaternionic spinorial formalism, which enable us to treat other signatures besides Lorentzian. Then, in Sec. \ref{Sec.LieAlgebra} we obtain the Lie algebra of $SL(2;\mathbb{H})$ and make a connection with the algebra of $SO(5,1)$, which enables us to give a physical interpretation for the transformations of the spin group. In Sec. \ref{Sec.MapTilde}, we introduce a map that connects two possible ways of representing a six-dimensional vector. The same map also connects two equivalent ways of representing a bivector and is essential to define the representation of a 3-vector, which is tackled in Sec. \ref{Sec.Bivec3Vec}. In Sec. \ref{Sec.ReviewSU4}, we review the basics of the four-component spinorial formalism. Based on the latter formalism we conclude that given a pair of spinors one must be able to construct a null vector, which proved to be the case in Sec. \ref{Sec.VectorFromSpinors}. Finally, in Sec. \ref{Sec.Bridge} we show how the objects in the two-component spinorial formalism are related their correspondents in the four-component formalism.

\section{Quaternions and the group $SL(2,\mathbb{H})$}\label{Sec.Quat}

The quaternions are an algebra that is denoted by $\mathbb{H}$ and that is generated by the basis $\{\bl{q}_{\mf{m}}\}=\{\bl{I},\bl{J},\bl{K},\bl{1}\}$, where
the indices $\mf{m},\mf{n},\mf{p},\mf{q}$ range from 1 to 4. Thus, a general quaternion $\bl{a}\in \mathbb{H}$ is written as
\begin{equation}\label{ijk}
 \boldsymbol{a} = a^{\mf{m}}\,\bl{q}_{\mf{m}} = a^1\,\bl{I} +  a^2\,\bl{J} +  a^3\,\bl{K} +  a^4 \,,
 \end{equation}
where $a^{\mf{m}}$ are real numbers.
The quaternionic conjugation is then denoted by setting a bar over the quaternion and is defined by
\begin{equation*}\label{QuatConj}
  \bar{\boldsymbol{a}} = -a^1\,\bl{I} -  a^2\,\bl{J} -  a^3\,\bl{K} +  a^4 \,.
\end{equation*}
The real part of the quaternion is then the one that is invariant by the action of the quaternionic conjugation, while the purely quaternionic part is the one that flips the sign under this operation. Thus, we can say that $a^4$ is the real part of the quaternion $\bl{a}$, whereas $\bl{b}=b^1\bl{I}+b^2\bl{J}+b^3\bl{K}$ is said to be a purely quaternionic number, since $\bar{\bl{b}}=-\bl{b}$. The product of two general quaternions can be obtained from the distributivity and associativity property of quaternions along with following product rules:
\begin{equation}\label{Algebra}
  \bl{I}^2 = \bl{J}^2 = \bl{K}^2 = \bl{I}\bl{J}\bl{K} = \bl{J}\bl{K}\bl{I} = \bl{K}\bl{I}\bl{J} = - 1  \,.
\end{equation}
In particular, one can use the latter relations to prove that $\bl{I}\bl{J}= - \bl{J}\bl{I} = \bl{K}$. Thus, the algebra is noncommutative. Hence, the order on the product of quaternions matter, so that we should be careful in our manipulations. Using the products (\ref{Algebra}) one can easily prove that the quaternionic conjugation obeys the following property
$$  \ovl{( \bl{a} \bl{b})} = \bar{\bl{b}}\,\bar{\bl{a}}\,, $$
for arbitrary $\bl{a},\bl{b}\in \mathbb{H}$. The modulus of a quaternion is then defined by
$$ |\bl{a}| = \sqrt{ \bl{a}\,\bar{\bl{a}}} = \sqrt{ \bar{\bl{a}} \, \bl{a}} = \sqrt{ (a^1)^2 + (a^2)^2 + (a^3)^2 + (a^4)^2 } \,.  $$
In particular, note that $\bl{a}$ always commute with its conjugate $\bar{\bl{a}}$ and that $ \bl{a}\,\bar{\bl{a}}$ commutes with every quaternion, since it is real. Note also that $|\bl{a} \bl{b}| = |\bl{a}||\bl{b}|$ and that a nonzero quaternion, $\bl{a}\neq 0$, always has an inverse, which is given by
$$ \bl{a}^{-1} = \frac{1}{|\bl{a}|^2}\,\bar{\bl{a}}\,, $$
which is such that $\bl{a}\bl{a}^{-1} = \bl{a}^{-1} \bl{a} =1$.


Now, let us consider the vector space of $2\times2$ quaternionic matrices, which is denoted henceforth by $M(2; \mathbb{H})$. In other words, a general element of $M(2; \mathbb{H})$ is given by
$$ A = \left[
         \begin{array}{cc}
            {A}_1^{\;\;1} &  {A}_1^{\;\;2} \\
            {A}_2^{\;\;1} &  {A}_2^{\;\;2} \\
         \end{array}
       \right]\,,
 $$
where $A_\alpha^{\;\;\beta}\in \mathbb{H}$. From now on we are assuming that Greek indices from the beginning of the alphabet, like $\alpha,\beta$, and $\gamma$ range from 1 to 2. The sum and product of two matrices are defined as usual. One can also define multiplication of a matrix by a scalar, but we must split this operation in two types, left multiplication and right multiplication, since quaternions do not commute. Summing up, if $\bl{a}\in \mathbb{H}$ and $A,B\in M(2; \mathbb{H})$ then we have the following usual operations:
$$ [A+B]_\alpha^{\;\;\beta} = A_\alpha^{\;\;\beta} + B_\alpha^{\;\;\beta},\;\;
[AB]_\alpha^{\;\;\beta} =  {A}_\alpha^{\;\;\gamma} {B}_\gamma^{\;\;\beta} ,\;\;
[\bl{a}A]_\alpha^{\;\;\beta} =\bl{a}\, A_\alpha^{\;\;\beta} ,\;\;
[A \bl{a}]_\alpha^{\;\;\beta} = A_\alpha^{\;\;\beta}\,\bl{a} .$$
One can also define the quaternionic conjugate of a matrix $A$ as being the matrix $\bar{A}$ whose components are the quaternionic conjugates of the components of $A$, namely $[\bar{A}]_\alpha^{\;\;\beta} = \ovl{A_\alpha^{\;\;\beta}}$. Particularly, matrix multiplication is distributive and associative. Transposition is defined as usual and denoted by a superscript $t$. Finally, we define the quaternionic hermitian conjugate of a matrix as the composition of the quaternionic conjugate followed by transposition and denoted by the superscript $\dagger$, so that
$$ [A^\dagger]^\alpha_{\;\;\beta} = \ovl{A_\beta^{\;\;\alpha}} \,.  $$
It turns out that several identities valid for complex matrices are also valid when quaternionic matrices are used \cite{Zhang}, as exemplified by
\begin{equation}\label{property}
  (A^{\dagger})^{-1} =  (A^{-1})^{\dagger}\,,\;\;
 (AB)^{\dagger} = B^{\dagger}A^{\dagger}  \,,\;\;
 (AB)^{-1} = B^{-1}A^{-1} \,,\;\;
(\bar{A})^{t} =  \overline{(A^{t})} \,.
\end{equation}
However, some other usual identities do not hold in the quaternionic context. For instance, in general it follows that
$$ (AB)^{t} \neq B^{t}A^{t} \,,\;\;
(\bar{A})^{-1} \neq \overline{(A^{-1})}  \,,\;\;
(A^{t})^{-1} \neq (A^{-1})^{t}  \,,\;\;
\overline{(AB)} \neq \bar{A}\,\bar{B}  \,.$$
Concerning the last inequality, one should also keep in mind that $\overline{(AB)} \neq \bar{B}\,\bar{A}$.


Defining the determinant of a quaternionic matrix can also be tricky.  It turns out that there exists a unique generalization of the determinant functional, here denoted by $\Delta$, which obeys the property $ \Delta(AB) = \Delta(A) \Delta(B), \, A,B \in M(2; \mathbb{H})$, called Dieudonn\'{e} determinant \cite{Dieu}. For the matrix
\begin{equation*}
A = \begin{bmatrix}
\boldsymbol{a} & \boldsymbol{b}\\
\boldsymbol{c} & \boldsymbol{d}
\end{bmatrix} \quad \text{with} \quad \boldsymbol{a},\,\boldsymbol{b},\,\boldsymbol{c},\, \boldsymbol{d} \in \mathbb{H} \,,
\end{equation*}
this determinant is defined as
\begin{equation}\label{det}
\Delta(A) = |\boldsymbol{a}\boldsymbol{d} - \boldsymbol{a}\boldsymbol{c}\boldsymbol{a}^{-1}\boldsymbol{b}|
=  |\bl{d}\boldsymbol{a} - \bl{d}\boldsymbol{b}\boldsymbol{d}^{-1}\boldsymbol{c}|
= | \bl{b}\boldsymbol{d}\boldsymbol{b}^{-1}\boldsymbol{a} - \bl{b}\boldsymbol{c} |
= | \bl{c}\boldsymbol{a}\boldsymbol{c}^{-1}\boldsymbol{d} - \bl{c}\boldsymbol{b} | \,.
\end{equation}
Note that, due to the modulus, this expression does not reduce to the determinant of a complex matrix if the entries are all complex. More precisely, if we assume that the components of the matrix do not have the parts $\bl{J}$ and $\bl{K}$, and if we imagine $\bl{I}$ as the complex imaginary unity, it follows that the Dieudonn\'{e} determinant does not become the usual determinant, rather we would have $\Delta(A) = |\textrm{det}(A)|$.

The inverse of the matrix $A$ is given by
\begin{equation*}
  A^{-1} = \left[
             \begin{array}{cc}
               (\boldsymbol{a} - \boldsymbol{b}\boldsymbol{d}^{-1}\boldsymbol{c})^{-1} & (\boldsymbol{c} - \boldsymbol{d}\boldsymbol{b}^{-1}\boldsymbol{a})^{-1} \\
               (\boldsymbol{b} - \boldsymbol{a}\boldsymbol{c}^{-1}\boldsymbol{d})^{-1} & (\boldsymbol{d} - \boldsymbol{c}\boldsymbol{a}^{-1}\boldsymbol{b})^{-1} \\
             \end{array}
           \right]\,.
\end{equation*}
Using the latter expression, one can easily us check that the inverse of a quaternionic matrix exists if, and only if, its  Dieudonn\'{e} determinant is different from zero. Indeed, it turns out that the Dieudonn\'{e} determinant vanishes if, and only if, a component of $A^{-1}$ diverges. For instance, $(A^{-1})_1^{\;\;1}$ is divergent if $\boldsymbol{a} - \boldsymbol{b}\boldsymbol{d}^{-1}\boldsymbol{c}=0$, which, due to Eq. (\ref{det}), imply a vanishing Dieudonn\'{e} determinant. Note that the above formula for the inverse matrix cannot be directly applied in the particular case in which one of the components of $A$ vanish. For instance, if $\bl{d}=0$ then the expression for $(A^{-1})_1^{\;\;1}$ cannot be computed due to the presence of $\bl{d}^{-1}$. However, taking the limit $\bl{d}\rightarrow 0$ in the expression $(A^{-1})_1^{\;\;1} = (\boldsymbol{a} - \boldsymbol{b}\boldsymbol{d}^{-1}\boldsymbol{c})^{-1}$ one obtain $(A^{-1})_1^{\;\;1} =0$, which is the correct choice.

Suppose that a quaternionic matrix $A_{\alpha}^{\;\;\beta}$ can be written as the product of two ``vectors'', namely $A_{\alpha}^{\;\;\beta}= \xi_\alpha\,\chi^\beta $. Then, it is simple matter to prove that its determinant vanish, $\Delta(A)=0$. Conversely, if the determinant of a quaternionic matrix vanish then it can be written as a product of ``vectors''. Indeed, assuming $A\neq 0$ then at least one of its components must be different from zero. Let us suppose $\bl{a}\neq 0$, so that $\bl{a}$ is invertible. Then, if $\Delta(A)=0$ it follows that $\boldsymbol{a}\boldsymbol{d} - \boldsymbol{a}\boldsymbol{c}\boldsymbol{a}^{-1}\boldsymbol{b}$ vanishes, which, in turn, implies that
$\boldsymbol{d} =  \boldsymbol{c}\boldsymbol{a}^{-1}\boldsymbol{b}$, so that the matrix is written as
$$ A = \left[
         \begin{array}{cc}
           \bl{a} & \bl{b} \\
           \bl{c} & \boldsymbol{c}\boldsymbol{a}^{-1}\boldsymbol{b} \\
         \end{array}
       \right] = \xi_\alpha\,\chi^\beta \;,\; \text{ with } \;\;
\xi_\alpha =\left[
              \begin{array}{c}
                1 \\
                \bl{c\,a}^{-1} \\
              \end{array}
            \right] \;\; \text{and} \;\;
\chi^\alpha =\left[
              \begin{array}{c}
                \bl{a} \\
                \bl{b} \\
              \end{array}
            \right] \,.
 $$
Thus, just as holds for $2\times 2$ complex matrices, a quaternionic $2\times 2$ matrix has vanishing determinant if, and only if, it can be written as a product of two column vectors. This result will be of relevance to establish that a $SO(5,1)$ vector is light-like if, and only if, its spinorial representation is the direct product of two spinors.

We are now able to define the $SL(2; \mathbb{H})$ group.  The special  linear  quaternionic  group is comprised of
the  $2 \times 2$  quaternionic  matrices  $Q\in M(2; \mathbb{H})$ satisfying  the  constraint  $\Delta(Q) =  1$,  where  the determinant is  defined  as shown in Eq. \eqref{det}. Since a general element of $ M(2; \mathbb{H})$ has 16 real degrees of freedom, and since the equation $\Delta(Q) =  1$ constrains one real degree of freedom, it follows that the group  $SL(2; \mathbb{H})$ has dimension 15. This is the same dimension of the orthogonal groups in six dimensions. Indeed, in what follows we shall prove that $SL(2; \mathbb{H})$ is deeply related to the Lorentz group $SO(5,1)$. More precisely, the former group is a double cover for the latter, i.e. the groups are homeomorphic and for each element of $SO(5,1)$ we can associate two elements in
$SL(2; \mathbb{H})$.

\section{Vectors of $\mathbb{R}^{5,1}$ in the quaternionic formalism}\label{Sec.Vectors}

In this section we shall see that the vectors of the Minkowski space in six dimensions, $\mathbb{R}^{5,1}$, can be represented by $2\times 2$ quaternionic matrices that are hermitian \cite{Kugo, Bengtsson88, Sudbery84}. In order to make the correspondence explicit, we introduce the following set of six quaternionic matrices:
\begin{equation}\label{qm}
  \sigma_{0} = \begin{bmatrix}
1 & 0\\
0 & 1
\end{bmatrix} \quad, \quad
\sigma_{\mf{m}} = \begin{bmatrix}
0 & \ovl{\bl{q}_{\mf{m}}}\\
\bl{q}_{\mf{m}} & 0
\end{bmatrix} \quad, \quad
\sigma_{5} = \begin{bmatrix}
1 & 0\\
0 & -1
\end{bmatrix} \,.
\end{equation}
This set is compactly referred to as $\{\sigma_\mu\}$. Here we are adopting the convention that the Greek indices from the middle of the alphabet, like $\lambda,\mu,\nu$ range from 0 to 5, comprising a total of six indices. These indices are identified as vector indices of Minkowski spacetime.  Using the Minkowski metric $\eta^{\mu\nu}=\textrm{diag}(-1,1,1,1,1,1)$ one can raise the vector index in this set of matrices and define another set given by
\begin{equation}\label{sigmatilde}
  \widetilde{\sigma}_\mu = \sigma^\mu = (-\sigma_0, \sigma_{\mf{m}}, \sigma_5)\,.
\end{equation}
These two sets of matrices satisfy the following algebra:
\begin{equation}\label{CA}
\sigma_{\mu}\,\widetilde{\sigma}_{\nu} + \sigma_{\nu}\,\widetilde{\sigma}_{\mu} = 2\,\eta_{\mu\nu} \mathbb{I}_2
\quad\text{and}\quad
 \widetilde{\sigma}_{\mu}\,\sigma_{\nu} + \widetilde{\sigma}_{\nu}\, \sigma_{\mu}\, = 2\,\eta_{\mu\nu} \mathbb{I}_2\,,
\end{equation}
where $\mathbb{I}_n$ is the $n\times n$ identity matrix. Thus, defining the $4\times 4$ gamma matrices as
\begin{equation}\label{Gamma1}
  \gamma_\mu = \left[
                  \begin{array}{cc}
                    0 & \sigma_\mu \\
                    \widetilde{\sigma}_\mu & 0 \\
                  \end{array}
                \right]
\end{equation}
it follows that the relation $\gamma_{(\mu} \gamma_{\nu)} = \eta_{\mu\nu}\mathbb{I}_4 $  holds, which is the  well known Clifford algebra. The $2\times 2$ matrices $\sigma_\mu$ and $\widetilde{\sigma}_\mu$ act on chiral spinors, while the $4\times 4$ matrices $\gamma_\mu$ act on  Dirac spinors, as we shall explain in the sequel.

Now, given a vector of components $V^\mu$ on $\mathbb{R}^{5,1}$, we can associate to it two $2\times 2$ quaternionic matrices, denoted here by $V$ and $\widetilde{V}$, that are defined by
\begin{equation*}
  V = V^\mu \,\sigma_\mu  \quad \text{ and } \quad \widetilde{V} = V^{\mu}\,\widetilde{\sigma}_{\mu}\,.
\end{equation*}
Since the matrices $\sigma_\mu$ and $\widetilde{\sigma}_{\mu}$ are all hermitian, i.e. they obey the relation $\sigma_\mu^\dagger = \sigma_\mu$ and
$\widetilde{\sigma}_{\mu}^\dagger = \widetilde{\sigma}_{\mu}$, it follows that vectors of $\mathbb{R}^{5,1}$ are represented by $2\times 2$ hermitian quaternionic matrices in the present formalism. Note that the number of degrees of freedom of a general hermitian matrix of $M(2;\mathbb{H})$ is six. Thus the sets $\{\sigma_\mu\}$ and $\{\widetilde{\sigma}_{\mu}\}$ are both frames for the subspace of hermitian matrices. In particular, this implies that, conversely, given a hermitian matrix of $M(2;\mathbb{H})$ we can associate to it a unique vector of $\mathbb{R}^{5,1}$, as long as we predefine which frame of hermitian matrices we are working with. More precisely, given $V$ or $\widetilde{V}$ the components of the vector are given by:
\begin{equation}\label{Vmu}
  V^\mu = \frac{1}{2}\,\textrm{Tr}[V\sigma_\mu] = \frac{1}{2}\,\textrm{Tr}[\widetilde{V} \widetilde{\sigma}_\mu] \quad
  \text{and} \quad V_\mu = \frac{1}{2}\,\textrm{Tr}[V\widetilde{\sigma}_\mu] = \frac{1}{2}\,\textrm{Tr}[\widetilde{V} \sigma_\mu]\,,
\end{equation}
where $\textrm{Tr}[A] = A_1^{\;\;1} + A_2^{\;\;2}$ denotes the usual trace of $A\in M(2;\mathbb{H})$. Thus, the association between vectors of $\mathbb{R}^{5,1}$ and hermitian matrices of $M(2;\mathbb{H})$ is one-to-one.

Nevertheless, up to now, this relation is solely based on the number of degrees of freedom, so that any six-dimensional vector space could be associated to $2\times 2$ hermitian quaternionic matrices, irrespective of the signature of the metric on the space, and even if the vector space is not endowed with a metric. However, from Eq. (\ref{Vmu}) one easily concludes that given two vectors $V$ and $W$ then it follows that
$$  \textrm{Tr}[V\widetilde{W}]  = \textrm{Tr}[\widetilde{V}W] =  2\,\eta_{\mu\nu}V^\mu W^\nu \,, $$
which hints that there exists a deeper connection with Lorentzian signature, since the Minkowski metric appears naturally. Moreover, computing the  Dieudonn\'{e} determinant of $V$ and $\widetilde{V}$ we can check that
\begin{equation}\label{DeltaVNorm}
\Delta(V)= \Delta(\widetilde{V}) =  |\eta_{\mu\nu}V^{\mu}V^{\nu}| \,,
\end{equation}
which gives another connection with the Lorentzian signature. Note also that for hermitian matrices, like $V$, when computing the Dieudonon\'{e} determinant it is not necessary to bother about non-commutativity of the quaternionic entries on the matrix, since the diagonal elements of a hermitian matrix are real and the non-diagonal ones are conjugated to each other and, therefore, commute. Hence, when dealing with hermitian quaternionic matrices, it is reasonable to use the usual expression for the determinant, which is given by
$$ \textrm{det}(V)  =  \textrm{det}\begin{bmatrix}
\ell & \overline{\boldsymbol{v}}\\
\boldsymbol{v} & n
\end{bmatrix}  = \ell n - \boldsymbol{v}\overline{\boldsymbol{v}} = - \eta_{\mu\nu}V^\mu V^\nu\,,  $$
where it is being assumed that $\eta_{\mu\nu}=\text{diag}(-,+,+,+,+,+)$. Note also that $\textrm{det}(V) = \textrm{det}(\widetilde{V})$. The advantage of using this determinant is that it captures the sign of the inner product $ V^\mu V_\mu$, whereas the Dieudonon\'{e} determinant  is insensitive to this sign. Indeed, if $V$ is an hermitian matrix then the relation between the two determinants is:
\begin{equation}\label{DETdet}
   \Delta(V) = |\textrm{det}(V)|\,.
\end{equation}
Note that the latter relation is not valid if $V$ is not hermitian. Thus, if an operation preserves the norm of the vector $V^\mu$ then it preserves the determinant of the matrices $V$ and $\widetilde{V}$. Conversely, if the determinant of $V$ is preserved by some operation then the norm of $V^\mu$ is also preserved. Since, by definition, a Lorentz transformation is a linear transformation on the vectors of the Minkowski space that preserve their norms,
it follows that we have obtained a relation between Lorentz transformations and operations that act on the matrices $V$ preserving their determinant. However, it is worth noting that since Lorentz transformations map vectors into vectors and since vectors are represented by hermitian matrices, it follows that the corresponding transformation in the space $M(2;\mathbb{H})$ must map hermitian matrices into hermitian matrices.

To each $Q \in SL(2; \mathbb{H})$, we define the action of $SL(2; \mathbb{H})$ on the space of hermitian matrices as follows:
\begin{equation}\label{Qaction}
V \overset{Q}{\longrightarrow} V'= QVQ^{\dagger}\,.
\end{equation}
With the latter definition it follows that the Hermitian nature of $V$ is preserved, i.e. if $V^\dagger = V$ then $V'^\dagger = V'$. This means that the latter action maps vectors into vectors, just as Lorentz transformations. Moreover,  the Dieudonon\'{e} determinant of $V$ does not change with the transformation, since this determinant obeys the property $\Delta(ABC ) = \Delta(A) \Delta(B) \Delta(C)$ for any $A,B,C\in M(2;\mathbb{H})$. Thus,
$\Delta(V)= \Delta(V')$ for any $Q\in  SL(2; \mathbb{H})$, i.e. for any $Q$ such that $\Delta(Q) = 1$. Since $\Delta(V)$ is the modulus of the norm of $V^\mu$ this implies that the transformations (\ref{Qaction}) preserve the norm of $V^\mu$.  Actually, since the Dieudonon\'{e} determinant is insensitive to a global sign, it follows that $V^\mu V_\mu$ could, in principle, flip the sign after the transformation. However, this could not happen for transformations connected to the identity. Moreover, after some algebra, it can be proved that if $V$ is hermitian then the following identity holds:
\begin{equation}\label{detQVQ}
  \textrm{det}(Q V Q^\dagger) = \textrm{det}(V) \,\Delta(Q)^2 \;,\;\;\forall\; Q\in M(2,\mathbb{H})\,.
\end{equation}
Thus, since we are assuming that $Q\in SL(2,\mathbb{H})$, it follows from the above identity that
$$V'^\mu V'_\mu = -\textrm{det}( V' ) = -\textrm{det}(Q V Q^\dagger) = -\textrm{det}(V) \,1^2 =   V^\mu V_\mu\,.   $$
This, along with the fact that the action \eqref{Qaction} is linear in $V$ implies that these transformations are contained in $SO(5,1)$. In order to prove that the whole $SO(5,1)$ is comprised of these transformations we just need to recall that dimension of the Lie group $SL(2,\mathbb{H})$ is 15, which is the dimension of the Lorentz group in six dimensions, namely $SO(5,1)$.


It can  be directly verified that for an arbitrary vector $V^\mu$ that the representations $V$ and $\widetilde{V}$ are related to each other by the following expression
\begin{equation}\label{VVtil}
\widetilde{V} = -\textrm{det}(\textrm{V})\, V^{-1}  \,.
\end{equation}
In particular, one can check that this is true for the frame $\sigma_\mu$.
Now, taking the inverse of both sides of \eqref{Qaction}, we get
\begin{equation}\label{Qactioni}
V^{-1} \overset{Q}{\longrightarrow} V'^{-1} = (Q^{\dagger -1}) V^{-1}Q^{-1} \,.
\end{equation}
Thus, using Eqs. (\ref{VVtil}) and (\ref{Qactioni}) along with the fact that $\textrm{det}(V) =  \textrm{det}(V')$ lead us to the following transformation rule for $\widetilde{V}$:
\begin{equation}\label{Qaction1}
\widetilde{V} \overset{Q}{\longrightarrow}  \widetilde{V}' = (Q^{\dagger -1})\, \widetilde{V}\, Q^{-1}\,.
\end{equation}
Thus, $V$ and $\widetilde{V}$ transform differently under the action of $SL(2; \mathbb{H})$. They are two different ways of representing a vector in the present formalism. We say that  $V$ and $\widetilde{V}$ carry different representations of the group $SL(2; \mathbb{H})$. As a final comment, note that the transformation (\ref{Qaction1}) also preserves the hermitian nature of $\widetilde{V}$ as well as the value of $\text{det}(\widetilde{V})$, so that it is a Lorentz transformation.

\section{Spinors in the quaternionic formalism}\label{Sec.SpinorRep}

The action of $Q \in SL(2; \mathbb{H})$ on a vector $V$ is quadratic on $Q$, so that it is not faithful. For instance, $Q$ and $-Q$ yield the same transformation on $V$, this is the reason why $SL(2; \mathbb{H})$ is said to be the double cover of $SO(5,1)$. However, there exist more fundamental objects that transform linearly by the action of the spin group, i.e. by the action of $Q \in SL(2; \mathbb{H})$. These objects are called spinors. Since $Q$ are $2 \times 2$ quaternionic matrices, a $SL(2; \mathbb{H})$ chiral spinor  is  2-component quaternionic object  of the form
\begin{equation}
\bl{\Psi}_+ = \begin{bmatrix}
\,\,\bl{\psi_1}\\
\bl{\psi_2}
\end{bmatrix}  \quad , \quad \bl{\psi_1},\bl{\psi_2}\in \mathbb{H} \,,
\end{equation}
whose transformation rule under the action of $SL(2; \mathbb{H})$ is given by
\begin{equation}\label{Transf2}
  \bl{\Psi}_+ \overset{Q}{\longrightarrow}  \bl{\Psi}_+' =  Q\,\bl{\Psi}_+ \quad , \quad \forall \quad Q \in SL(2; \mathbb{H})\,.
\end{equation}
By the very definition of representation of a group, we say that $\bl{\Psi}_+$ carry a representation of $SL(2,\mathbb{H})$.
In terms of components, the latter transformation is given by
\begin{equation}\label{pdsd}
\boldsymbol{2} : \,\, \boldsymbol{\psi}_{\alpha} \overset{Q}{\longrightarrow} \bl{\psi}'_\alpha = Q_{\alpha}^{\phantom{\alpha}\beta}\,\boldsymbol{\psi}_{\beta} \quad , \quad \forall \quad Q \in SL(2; \mathbb{H}) \,,
\end{equation}
where we have started to employ the convention that the representation will be labelled by its dimension in boldface. Let us denote the quaternionic vector space spanned by the spinors that transform as in Eq. (\ref{Transf2}) by $\mathbb{S}$ (recall that quaternions do not form a field, but it is sensible to define a vector space over the quaternions). In other words, we say that the elements of $\mathbb{S}$ are the ones that carry the representation $\bl{2}$ of $SL(2; \mathbb{H})$.

On the dual of this space, $\mathbb{S}^*$, the group $SL(2; \mathbb{H})$ then acts in the inverse manner. We then say that the elements of $\mathbb{S}^*$ carry the representation  $\bl{2}^{-1}$ of $SL(2; \mathbb{H})$. The spinors that carry the representation $\bl{2}^{-1}$ are two-component row-vectors that transform as follows
$$ \bl{\Psi}_- \overset{Q}{\longrightarrow}  \bl{\Psi}_-' = \bl{\Psi}_-\,Q^{-1} \,,$$
where
$$ \bl{\Psi}_- = \big[\bl{\psi}^1\;\; \bl{\psi}^2\big] \quad , \quad \bl{\psi}^1,\bl{\psi}^2\in \mathbb{H}\,.$$
In terms of components we have
$$\bl{2}^{-1}:\;\; \bl{\psi}^\alpha  \overset{Q}{\longrightarrow} \bl{\psi}'^\alpha = \bl{\psi}^\beta \,(Q^{-1})_\beta^{\;\;\alpha} \,.$$
In the above definitions, the subscripts $\pm$ in $\bl{\Psi}_+$ and $\bl{\Psi}_-$ are there just to distinguish between column vectors (denoted by $+$) and row vectors (denoted by $-$). Note that the action of the group on $\bl{\Psi}_+$ is on the left, while the action of the group on objects like $\bl{\Psi}_-$ is on the right. It is important to keep this distinction in mind.
One should check that the latter objects, i.e. the elements of $\mathbb{S}^*$, also carry a representation of the $SL(2, \mathbb{H})$ group. Indeed, suppose that $P,Q,R$ are elements of $SL(2, \mathbb{H})$ such that $PQ=R$, i.e. $R$ is obtained by the action of $Q$ followed by the action of $P$. Then, when performing the transformations $Q$ and $P$ successively, we have
$$ \bl{\Psi}_- \overset{Q}{\longrightarrow}   \bl{\Psi}_-\,Q^{-1}  \overset{P}{\longrightarrow} \bl{\Psi}_-\,Q^{-1}\,P^{-1} = \bl{\Psi}_-\,(PQ)^{-1}  =  \bl{\Psi}_-\,R^{-1} \,$$
which is the desired action. Note that it is essential that the action of $Q^{-1}$ comes from the right in order to yield a representation. Moreover, note that a transformation law like $\bl{\Psi}_-\rightarrow \bl{\Psi}_-Q$ does not yield a representation. The left action of an element of $\mathbb{S}^*$ on an element of $\mathbb{S}$ is invariant under the spin group transformation. Indeed,
$$ \bl{\phi}'^\alpha \bl{\psi}'_\alpha = \bl{\Phi}_-' \, \bl{\Psi}_+' =   \bl{\Phi}_-\,Q^{-1} \, Q \,\bl{\Psi}_+ =  \bl{\Phi}_-\, \bl{\Psi}_+
= \bl{\phi}^\alpha \bl{\psi}_\alpha \,.  $$
Note, however, that $\bl{\psi}_\alpha \bl{\phi}^\alpha $ is not invariant in general, since quaternions do not commute.

Besides the latter linear representations of $SL(2; \mathbb{H})$, one can check that the left action of $Q^{\dagger -1}$ also forms a representation. Indeed, take $P, Q, R \in SL(2; \mathbb{H})$ such that
$PQ = R$, then by means of \eqref{property} we easily obtain that:
\begin{equation}
P^{\dagger -1}Q^{\dagger -1} = (Q^{\dagger}\,P^{\dagger})^{-1} = (PQ)^{\dagger -1} = R^{\dagger -1} \,.
\end{equation}
Moreover, it can be proved that $Q$ and $Q^{\dagger -1}$ are in fact independent representations of $SL(2; \mathbb{H})$. That is, there is no $2 \times 2$  invertible matrix $M$ such that $Q^{\dagger -1} = M Q M^{-1}$ for all $Q \in SL(2; \mathbb{H})$. Indeed, the latter condition can be equivalently written as:
$$ M = Q^\dagger\, M\, Q \;\;\forall\,Q\,\in\, SL(2; \mathbb{H})\,. $$
Thus, imposing such constraint for the matrices $Q=Q_1$ and $Q=Q_2$, which are defined by
$$ Q_1 \equiv \left[
                \begin{array}{cc}
                  1 & 1 \\
                  0 & 1 \\
                \end{array}
              \right] \;\; \text{and}\;\;
  Q_2 \equiv \left[
                \begin{array}{cc}
                  1 & 0 \\
                  1 & 1 \\
                \end{array}
              \right] \,,$$
and are both elements $SL(2; \mathbb{H})$, it follows that the conditions  $M = Q_1^\dagger M Q_1$ and $M = Q_2^\dagger M Q_2$ have as the only simultaneous solution $M=0$, which is nonsense, since $M$ must be invertible. Therefore, we conclude that there exists no invertible matrix $M$ such that the equation $Q^{\dagger -1} = M Q M^{-1}$ holds for an arbitrary $Q$ possessing unit Dieudonon\'{e} determinant. This proves that the representation given by the left action of $Q\in SL(2; \mathbb{H})$ is independent from the representation given by the left action of $(Q^\dagger)^{-1}$. The latter representation will be denoted here by $\bar{\bl{2}}^{-1}$. More explicitly, the objects carrying the latter representation are two-component quaternionic column-vectors $\bar{\bl{\Psi}}_+$ that transform as
$$  \bl{\bar{2}}^{-1} : \;\;  \bar{\bl{\Psi}}_+  \overset{Q}{\longrightarrow} \bar{\bl{\Psi}}_+' =  (Q^\dagger)^{-1} \, \bar{\bl{\Psi}}_+\,. $$
Likewise, if $\bar{\bl{\Psi}}_-$ is a two-component quaternionic row-vector, it follows that the following transformation also provides a representation for the group $SL(2, \mathbb{H})$:
$$ \bl{\bar{2}} : \;\; \bar{\bl{\Psi}}_- \overset{Q}{\longrightarrow} \bar{\bl{\Psi}}_-'=  \bar{\bl{\Psi}}_-\,Q^{\dagger} \,.$$
The objects $\bar{\bl{\Psi}}_-$ are said to span the quaternionic vector space $\bar{\mathbb{S}}$, whereas  the objects $\bar{\bl{\Psi}}_+$ span the dual space $\bar{\mathbb{S}}^*$.
In order to distinguish from the previous objects that transform under the representations $\bl{2}$ and $\bl{2}^{-1}$, we shall use a dot over the index that label the components of an object that transform under $\bl{\bar{2}}$ or $\bl{\bar{2}}^{-1}$. More explicitly, we have
$$  \bl{\bar{2}}^{-1}: \;\;  \bar{\bl{\Psi}}_+=\left[
                                               \begin{array}{c}
                                                 \bl{\psi}^{\dot{1}} \\
                                                 \bl{\psi}^{\dot{2}}  \\
                                               \end{array}
                                             \right]
                                             \quad \text{ and }
 \quad  \bl{\bar{2}} : \;\; \bar{\bl{\Psi}}_-= \big[ \bl{\psi}_{\dot{1}}\;\; \bl{\psi}_{\dot{2}}  \big]\,,  $$
with the transformation rules being
$$  \bl{\bar{2}}^{-1}: \;   \bl{\psi}^{\dot{\alpha}} \overset{Q}{\longrightarrow}   \bl{\psi}'^{\dot{\alpha}} =  (Q^{\dagger-1})^{\dot{\alpha}}_{\;\;\dot{\beta}} \, \bl{\psi}^{\dot{\beta}}
\quad \text{ and }
 \quad  \bl{\bar{2}} :\;
 \bl{\psi}_{\dot{\alpha}} \overset{Q}{\longrightarrow}   \bl{\psi}'_{\dot{\alpha}} =
  \bl{\psi}_{\dot{\beta}}\, (Q^{\dagger})^{\dot{\beta}}_{\;\;\dot{\alpha}} \,. $$
Note that the contraction $  \bl{\psi}_{\dot{\alpha}} \bl{\psi}^{\dot{\alpha}}  $ is invariant under the action of $SL(2, \mathbb{H})$, whereas
$  \bl{\psi}^{\dot{\alpha}}\bl{\psi}_{\dot{\alpha}}   $ is generally not. The reason why the order of up and down indices in $Q_{\alpha}^{\phantom{\alpha}\beta}$ and $(Q^{\dagger-1})^{\dot{\alpha}}_{\;\;\dot{\beta}}$ are different is because the operation $\dagger$ (hermitian conjugation) involves transposition, so that the upper index, which is the second index in $Q_{\alpha}^{\phantom{\alpha}\beta}$ becomes the first index after the hermitian conjugation.

It is important noting that the representation $\bl{\bar{2}}$ can be obtained from the representation $\bl{2}$ by hermitian conjugation. Indeed, taking the hermitian conjugation of the transformation law of the representation
$\bl{2}$, given in Eq. (\ref{Transf2}), we obtain
\begin{equation}\label{Transf3}
  \bl{\Psi}_+^\dagger \overset{Q}{\longrightarrow}  (\bl{\Psi}_+')^\dagger =  \,(\bl{\Psi}_+)^\dagger Q^\dagger \,,
\end{equation}
which is the transformation law of the representation $\bar{\bl{2}}$. Likewise,   $\bl{2}^{-1}$ and $\bl{\bar{2}}^{-1}$ are also related to each other by hermitian conjugation. Therefore, if $\psi_\alpha$ transforms on the representation $\bl{2}$ then its quaternionic conjugate, $\overline{\psi_\alpha}$ transforms on the representation $\bl{\bar{2}}$. In the same fashion, if $\phi^{\dot{\alpha}}$ transforms on the representation $\bl{\bar{2}}^{-1}$ then its quaternionic conjugate $\overline{\phi^{\dot{\alpha}}}$ will carry the representation $\bl{2}^{-1}$. Summing up:
\begin{equation}\label{BarDagger}
  \begin{array}{ll}
      \bl{2} &\overset{\dagger}{\longrightarrow} \,\,\,\,\, \bl{\bar{2}} \\
      \psi_\alpha &\overset{\dagger}{\longrightarrow} \,\,\,\,\,\bar{\psi}_{\dot{\alpha}} \equiv \overline{(\psi_\alpha)}
    \end{array}\quad \quad, \quad \quad
     \begin{array}{ll}
      \bl{2}^{-1} &\overset{\dagger}{\longrightarrow} \,\,\,\,\,\bl{\bar{2}}^{-1} \\
      \psi^\alpha &\overset{\dagger}{\longrightarrow}  \,\,\,\,\,\bar{\psi}^{\dot{\alpha}} \equiv \overline{(\psi^\alpha)}
    \end{array}\,.
\end{equation}
Although from the component point of view the transformation $\bl{2} \rightarrow \bl{\bar{2}}$ amounts to just taking the quaternionic conjugate, it is worth recalling that these representations have a fundamental difference between them, namely the action of the group on the objects that carry the representation $\bl{2}$ is on the left, while for the spinors that carry the representation $\bl{\bar{2}}$ the action is on the right.


\section{Building higher rank representations from the spinor representations}\label{Sec.Tensors}

The objects that transform under the spin group $SL(2;\mathbb{H})$ according to the representations $\bl{2}$, $\bl{2}^{-1}$, $\bar{\bl{2}}$, and  $\bar{\bl{2}}^{-1}$ are all called spinors. Since they carry the lower-dimensional faithful representations of $SL(2;\mathbb{H})$ we say that they are on the fundamental representations of this group. One could also say that $\bl{2}$ alone is the fundamental representation, since with it one can generate the other ones. Indeed, $\bar{\bl{2}}$ can be seen as a byproduct of $\bl{2}$ through the use of hermitian conjugation, while the elements carrying the representation $\bl{2}^{-1}$ are the ones that live in the dual space of the vector space formed by the elements transforming according to $\bl{2}$. Likewise, $\bar{\bl{2}}^{-1}$ can be obtained from $\bl{2}$ by taking the hermitian conjugate followed by inverse operation. These spinors representations are the building blocks of any other representation of $SL(2;\mathbb{H})$.

For instance, if $\xi_\alpha$ carries the representation $\bl{2}$ while $\chi_{\dot{\alpha}}$ carries the representation $\bar{\bl{2}}$ then the object
$T_{\alpha \dot{\beta}} \equiv \xi_\alpha \chi_{\dot{\alpha}}$ is said to carry the representation $\bl{2}\otimes \bar{\bl{2}}$ and transforms as follows:
$$ T_{\alpha \dot{\beta}}  \overset{Q}{\longrightarrow} Q_\alpha^{\;\;\gamma}\,T_{\gamma \dot{\delta}}\, (Q^\dagger)^{\dot{\delta}}_{\;\;\dot{\beta}} \,. $$
Assuming that $T$ is the matrix with entries $ T_{\alpha \dot{\beta}}$, it follows that the transformation of $T$ is $T\rightarrow Q T Q^\dagger$. This is the same transformation law of the hermitian matrix $V = V^\mu \sigma_\mu$ that represents the vector $V^\mu$. Therefore, we say that $V$ carries the representation $\bl{2}\otimes \bar{\bl{2}}$. In the same fashion, we have that $\widetilde{V} = V^\mu \widetilde{\sigma}_\mu$, which transforms as $\widetilde{V} \rightarrow Q^{\dagger -1}\widetilde{V} Q^{-1}$, is said to carry the representation $\bar{\bl{2}}^{-1}\otimes \bl{2}^{-1}$. Thus, in this spinorial formalism, a vector of $\mathbb{R}^{5,1}$ can be associated to objects that carry two different representations of $SL(2;\mathbb{H})$:
$$  V^\mu\;\rightarrow\; \left\{
            \begin{array}{ll}
              V_{\alpha\dot{\beta}} \; (\text{contained in rep. } \bl{2}\otimes \bar{\bl{2}}) \\
              \quad \\
              \widetilde{V}^{\dot{\alpha}\beta} \; (\text{contained in rep. } \bar{\bl{2}}^{-1}\otimes \bl{2}^{-1})
            \end{array}
          \right.
 $$
The type of indices that we associate to $V$ and $\widetilde{V}$, namely up or down indices and with or without dot, is very important, since it dictates the way these objects transform. From what has just been discussed, we conclude that the matrices $\sigma_\mu$ should be written in the index notation as
$(\sigma_\mu)_{\alpha\dot{\beta}}$, whereas $\widetilde{\sigma}_\mu$ should be written as $(\widetilde{\sigma}_\mu)^{\dot{\alpha}\beta}$.

In the present formalism, a Dirac spinor is a four-component column vector with quaternionic entries. Let us now obtain which representation of $SL(2;\mathbb{H})$ this object carries. In order to do so, we recall that according to the general formalism of Clifford algebra, the gamma matrices, here introduced in Eq. (\ref{Gamma1}), should map a Dirac spinor into a Dirac spinor. Therefore, since the gamma matrices adopted here have the form
$$ \gamma_\mu = \left[
                  \begin{array}{cc}
                    0 & (\sigma_\mu)_{\alpha\dot{\beta}} \\
                    (\widetilde{\sigma}_\mu)^{\dot{\beta}\alpha} & 0 \\
                  \end{array}
                \right]\,,
 $$
it follows that a Dirac spinor carries the representation $\bl{2}\oplus \bar{\bl{2}}^{-1}$, i.e. a Dirac spinor has the form
$$ \Phi = \left[
            \begin{array}{c}
              \bl{\phi}_\alpha \\
	              \bl{\phi}^{\dot{\beta}} \\
            \end{array}
          \right] \,,
 $$
so that the action of $\gamma_\mu$ in $\Phi$ yields
$$ \gamma_\mu \Phi =  \left[
                  \begin{array}{cc}
                    0 & (\sigma_\mu)_{\alpha\dot{\beta}} \\
                    (\widetilde{\sigma}_\mu)^{\dot{\beta}\alpha} & 0 \\
                  \end{array}
                \right] \left[
            \begin{array}{c}
              \bl{\phi}_\alpha \\
              \bl{\phi}^{\dot{\beta}} \\
            \end{array}
          \right]  =
\left[
            \begin{array}{c}
              (\sigma_\mu)_{\alpha\dot{\gamma}} \bl{\phi}^{\dot{\gamma}}  \\
              (\widetilde{\sigma}_\mu)^{\dot{\beta}\delta}  \bl{\phi}_\delta\\
            \end{array}
          \right]\,, $$
which is again an object on the $\bl{2}\oplus \bar{\bl{2}}^{-1}$ representation. Note that we have the same symbol, namely $\bl{\phi}$, to denote the two independent parts of the Dirac spinor, $\bl{\phi}_\alpha$ and $\bl{\phi}^{\dot{\alpha}}$. This is reasonable because there is no canonical map connecting the representations $\bl{2}$ and $\bar{\bl{2}}^{-1}$. Therefore, the object $\bl{\phi}^{\dot{\alpha}}$ cannot be seen as a transformed version of the object $\bl{\phi}_\alpha$, differently from what happens with $V^\mu$ and $V_\mu$ which carry the same degrees of freedom of a single object.

The so-called chirality matrix is defined by
$$ \Gamma = -\gamma_0\gamma_1 \gamma_2\gamma_3\gamma_4\gamma_5 $$
and can be computed by means of Eq. (\ref{Gamma1}) along with the definitions of $\sigma_\mu$ and $\widetilde{\sigma}_\mu$, which eventually yields
\begin{equation}\label{ChiralityMatrix}
  \Gamma = \left[
               \begin{array}{cc}
                 \mathbb{I}_2 & 0 \\
                 0 & -\mathbb{I}_2 \\
               \end{array}
             \right]\,.
\end{equation}
The so-called chiral spinors are the Dirac spinors that are eigenvectors of the chirality matrix $\Gamma$. From Eq. (\ref{ChiralityMatrix}), we can see that the eigenvalues of $\Gamma$ are $\pm 1$. Eigenvectors with eigenvalue $+1$ are said to have positive chirality, while those with eigenvalue $-1$ have negative chirality. Taking a look at Eq. (\ref{ChiralityMatrix}), we see that Dirac spinors of the form
$$ \Phi^+ = \left[
            \begin{array}{c}
              \bl{\phi}_\alpha \\
              0 \\
            \end{array}
          \right]  \quad \text{ and } \quad \Phi^- = \left[
            \begin{array}{c}
              0 \\
              \bl{\phi}^{\dot{\beta}} \\
            \end{array}
          \right]   $$
are chiral spinors of positive and negative chirality respectively. Thus, for instance, the chiral spinor $\Phi^+$ has only the component on the representation $\bl{2}$, with the other component vanishing. This is why we can say that spinors on the representation $\bl{2}$ are chiral spinors of positive chirality, whereas spinors on the representation $\bar{\bl{2}}^{-1}$ are chiral spinors of negative chirality.

We have already used the spinorial representations to build higher order representations like $\bl{2}\otimes \bar{\bl{2}}$ and $\bl{2}\oplus \bar{\bl{2}}^{-1}$. Let us now discuss about more general representations that can be built. Concerning direct sum representations like $\bl{2}\oplus \bar{\bl{2}}^{-1}$ and $\bl{2}\oplus \bar{\bl{2}}$, for example, there is not much to be commented, since one can easily check that the direct sum of representations yield a representation. On the other hand, regarding tensor representations, like $\bl{2}\otimes \bar{\bl{2}}$, we must be more careful. When we are dealing with representations over the real or the complex field it is true that the direct product of two representations yield a representation. However, the non commutativity of quaternions makes things much more tricky. As we shall see, only certain tensors will carry a representation of the group.
Suppose, for instance, that the vectors $Z^a$ of an $n-$dimensional vector space over the complex field carry a representation $\bl{n}$ of some group whose elements are generically denoted by $L^a_{\;\;b}$. So, they transform as
$$ Z^a \rightarrow Z'^a = L^a_{\;\;b}Z^b \,. $$
Then, the tensor representation $\bl{n}\otimes \bl{n}$ is carried by an object $T^{ab}$ which transforms as
\begin{equation}\label{TenRepQuat}
  T^{a_1a_2} \rightarrow T'^{a_1a_2} = L^{a_1}_{\;\;\;b_1}L^{a_2}_{\;\;\;b_2}\, T^{b_1b_2}\,.
\end{equation}
In the latter expression it does not matter the order of the product $L^{a_1}_{\;\;\;b_1}L^{a_2}_{\;\;\;b_2}$, we could equivalently have written $L^{a_2}_{\;\;\;b_2}L^{a_1}_{\;\;\;b_1}$. Nevertheless, when the matrices $L^a_{\;\;b}$ are quaternionic the order is quite important. Therefore, when defining tensor representations of groups represented matrices with quaternionic entries we must establish a convention about the order. The adopted convention could be, for example, the one shown in Eq. (\ref{TenRepQuat}), i.e. the order of the transformation matrices  $L^a_{\;\;b}$ will follow the order of the indices of the tensor being transformed, and they come from the left (note that the tensor itself is also quaternionic in this scenario). However, it turns out that this does not lead to a representation of the group. Indeed, suppose we perform two transformations, first $L$ and then $M$, yielding the transformation $N = M\circ L$. Then, using the mentioned order convention we would have
\begin{equation}\label{TenRepQuat2}
  T^{ab} \overset{ M\circ L}{\longrightarrow}   M^{a}_{\;\;e}M^{b}_{\;\;f}\,\left(  L^{e}_{\;\;c}L^{f}_{\;\;d}\, T^{cd} \right) \;\neq\;
  M^{a}_{\;\;e}\,  L^{e}_{\;\;c} \,M^{b}_{\;\;\;f}\, L^{f}_{\;\;d}\, T^{cd} = N^{a}_{\;\;c}N^{b}_{\;\;d}\, T^{cd} \,.
\end{equation}
The inequality comes from the fact that the components $M^{b}_{\;\;f}$ and $ L^{e}_{\;\;c}$ are now quaternionic, so that they do not commute in general. Therefore, this tensorial transformation does not provide a representation for the group. Another possibility of transformation could be the following: \begin{equation}\label{TenRepQuat3}
  T^{ab} \overset{ L}{\longrightarrow}  L^{a}_{\;\;c}\, T^{cd}\,L^{b}_{\;\;d}\,.
\end{equation}
However, performing the transformations $L$ and $M$ successively we have
\begin{equation}\label{TenRepQuat4}
  T^{ab} \overset{ M\circ L}{\longrightarrow}   M^{a}_{\;\;e}\left(  L^{e}_{\;\;c}\, T^{cd}\,L^{f}_{\;\;d} \right)M^{b}_{\;\;f}
  =N^{a}_{\;\;c}\, T^{cd}\,L^{f}_{\;\;d} M^{b}_{\;\;f}
\end{equation}
which shows that the latter order convention, likewise, does not provide a tensor representation if $L^{f}_{\;\;d}$ doe not commute with $M^{b}_{\;\;f}$. On the other hand, suppose that the indices of $T^{ab}$ transform according to different representations, so that it is more suitable to write $T^{a\hat{b}}$. For instance, consider that the index $\hat{b}$ transforms according the matrix $\hat{L}_{\hat{a}}^{\;\;\hat{b}}$, where $\wedge$ is an operation such that $\widehat{M L} = \hat{L} \hat{M}$. For instance, $\wedge$ could be the inverse operation or the hermitian conjugate $\dagger$. We shall refer to this representation as $\hat{\bl{n}}$, whose transformation rule is
$$ Z^{\hat{a}} \rightarrow Z'^{\hat{a}} = Z^{\hat{b}}\,\hat{L}_{\hat{b}}^{\;\;\hat{a}} \,. $$
Then, a tensor with the index structure $T^{a\hat{b}}$ would transform as:
\begin{equation}\label{TenRepQuat5}
  T^{a\hat{b}} \overset{ L}{\longrightarrow}  L^{a}_{\;\;c}\, T^{c\hat{d}}\,\hat{L}_{\hat{d}}^{\;\;\hat{b}}\,.
\end{equation}
So, performing the transformations $L$ and $M$ successively, we are led to
\begin{align*}
 T^{a\hat{b}} \overset{ M\circ L}{\longrightarrow} \,\,  &M^{a}_{\;\;e}\left(  L^{e}_{\;\;c}\, T^{c\hat{d}}\,\hat{L}_{\hat{d}}^{\;\;\hat{f}} \right)\hat{M}_{\hat{f}}^{\;\;\hat{b}}
  =N^{a}_{\;\;c}\, T^{c\hat{d}}\,(\hat{L} \hat{M})_{\hat{d}}^{\;\;\hat{b}} =
  N^{a}_{\;\;c}\, T^{c\hat{d}}\,(\widehat{ML})_{\hat{d}}^{\;\;\hat{b}} \nonumber\\
&=   N^{a}_{\;\;c}\, T^{c\hat{d}}\,\hat{N}_{\hat{d}}^{\;\;\hat{b}}\,, \label{TenRepQuat6}
\end{align*}
which is what is expected from a true representation. Thus, we can say that $\bl{n}\otimes \hat{\bl{n}}$ is a representation of the quaternionic group. Hence, we conclude that in order to form a two index tensor representation we must pair a representation that naturally acts from the left ($\bl{n}$ in this example) with a representation that naturally acts from the right ($\hat{\bl{n}}$ in this example). Concerning our problem of interest, namely the representations of the group $SL(2;\mathbb{H})$, it follows that it is only reasonable to pair the following representations: $\bl{2}\otimes \bl{2}^{-1}$, $\bl{2}\otimes \bl{\bar{2}}$, $\bl{\bar{2}}^{-1}\otimes \bl{2}^{-1}$, and $\bl{\bar{2}}^{-1}\otimes \bl{\bar{2}}$ which lead to the following tensorial representations:
\begin{equation}\label{TensorRep}
   T_\alpha^{\;\;\beta} \;\;,\quad T_{\alpha \dot{\beta}} \;\;,\quad T^{\dot{\alpha}\beta} \;\;\text{and}\quad T^{\dot{\alpha}}_{\;\;\dot{\beta}} \,.
\end{equation}
On the other hand, objects with other index structures, like $T_{\alpha\beta}$ and $T_{\dot{\alpha} \dot{\beta}}$, do not carry representations of $SL(2;\mathbb{H})$.

It is also interesting noticing that, differently from what it would be expected if commuting numbers were used, the trace of the representation $T_\alpha^{\;\;\beta}$, namely $T_\alpha^{\;\;\alpha}$, is not invariant by the action of the group. Indeed, since this tensor transforms as
\begin{equation}\label{Ttransf}
 T_\alpha^{\;\;\beta} \overset{ Q}{\longrightarrow}
 T_\alpha^{'\;\;\beta} = Q_\alpha^{\;\;\sigma}\, T_\sigma^{\;\;\gamma}\,(Q^{-1})_{\gamma}^{\;\;\beta}\,,
 \end{equation}
it follows that its trace obeys the following transformation law
\begin{equation*}
 T_\alpha^{\;\;\alpha} \overset{ Q}{\longrightarrow}
 T_\alpha^{'\;\;\alpha} = Q_\alpha^{\;\;\sigma}\, T_\sigma^{\;\;\gamma}\,(Q^{-1})_{\gamma}^{\;\;\alpha}\,.
 \end{equation*}
Should we be dealing with commuting numbers, the right hand side could be written as
$(Q^{-1})_{\gamma}^{\;\;\alpha} Q_\alpha^{\;\;\sigma} T_\sigma^{\;\;\gamma} = \delta_\gamma^{\;\;\sigma}T_\sigma^{\;\;\gamma}$, so that the trace would be invariant. But this is not the case here, since quaternions do not commute. This is another example of how tricky quaternionic representations can be.
However, it turns out that the real part of the trace is preserved, i.e. if $T_\alpha^{\;\;\beta}$ obeys the transformation law (\ref{Ttransf}) then the combination $(T_\alpha^{\;\;\alpha}+\overline{T_\alpha^{\;\;\alpha}})$ is invariant by the action of any invertible quaternionic matrix
$Q\in GL(2;\mathbb{H})$. This can be seen from the fact that if $T_\alpha^{\;\;\beta} = \lambda \delta_\alpha^{\;\;\beta}$, with $\lambda = \overline{\lambda}$, then it is invariant by the action of $Q$, since $\lambda$ commutes with every element of the algebra. This means that the 16-dimensional representation $T_\alpha^{\;\;\beta}$ is not irreducible, it splits into a direct sum of a 15-dimensional representation, comprised of all tensors $\check{T}_\alpha^{\;\;\beta}$ whose real part of the trace vanish, plus a 1-dimensional representation formed by the tensors that are proportional to $\delta_\alpha^{\;\;\beta}$ with the proportionality factor being a real constant. In order to check this more explicitly, one can use the fact that a general element of $GL(2;\mathbb{H})$, the group of invertible matrices of $M(2;\mathbb{H})$, can be generated by the composition of the following transformations:
\begin{align*}
 R_1 = \left[
           \begin{array}{cc}
             r_1 & 0 \\
             0 & r_2 \\
           \end{array}
         \right]\,,&\;\;
R_2 = \left[
           \begin{array}{cc}
             \cosh r_3 & \sinh r_3 \\
             \sinh r_3 & \cosh r_3 \\
           \end{array}
         \right]\,, \;\;
R_3 = \left[
           \begin{array}{cc}
             \cos r_4 & \sin r_4 \\
             -\sin r_4 & \cos r_4 \\
           \end{array}
         \right]\,, \\
Q_1 =& \left[
           \begin{array}{cc}
             1 & 0 \\
             \omega_1 & 1 \\
           \end{array}
         \right]\,,\;\;
Q_2 = \left[
           \begin{array}{cc}
             1 & \omega_2 \\
             0 & 1 \\
           \end{array}
         \right]\,,\;\;
 Q_3 = \left[
           \begin{array}{cc}
             e^{\omega_3} & 0 \\
             0 & e^{\omega_4} \\
           \end{array}
         \right]\,,
\end{align*}
where $r_{\mf{m}}$ are arbitrary real numbers (with $r_1 r_2\neq 0$), whereas $\omega_{\mf{m}}$ are arbitrary purely quaternionic numbers, namely $\overline{\omega_{\mf{m}}} = - \omega_{\mf{m}}$. Of course, the trace of the tensor $T_\alpha^{\;\;\beta}$ is invariant under the action of $R_1$, $R_2$, and $R_3$, since these are real matrices. On the other hand, the matrices $Q_1$, $Q_2$, and $Q_3$ obey the property $Q = \overline{Q^{-1}}$, as can be easily checked. Thus, the transformation of the quaternionic conjugate of the matrix $T$ can be written as
$$  \overline{T_\alpha^{\;\;\beta}} \overset{ Q}{\longrightarrow}
 \overline{T_\alpha^{\;\;\beta}}^{\,'} = Q_\alpha^{\;\;\sigma}\, \overline{T_\sigma^{\;\;\gamma}}\,(Q^{-1})_{\gamma}^{\;\;\beta} \,.$$
Hence, the real part of the trace transforms as
\begin{equation}\label{Tracebar}
  \frac{1}{2}\left( T_\alpha^{\;\;\alpha} + \overline{T_\alpha^{\;\;\alpha}} \right) \overset{ Q}{\longrightarrow}
Q_\alpha^{\;\;\sigma}\,(Q^{-1})_{\gamma}^{\;\;\alpha}\left( T_\sigma^{\;\;\gamma} + \overline{T_\sigma^{\;\;\gamma}} \right) \,
\end{equation}
where it has been used that $\left( T_\sigma^{\;\;\gamma} + \overline{T_\sigma^{\;\;\gamma}} \right)$ are real numbers and, therefore, commute with $(Q^{-1})_{\gamma}^{\;\;\alpha}$. Finally, one can check that for the matrices $Q_1$, $Q_2$, and $Q_3$ it is true that $Q_\alpha^{\;\;\sigma}\,(Q^{-1})_{\gamma}^{\;\;\alpha} = \delta_\gamma^{\;\;\sigma}$, which along with Eq. (\ref{Tracebar}) implies that the real part of the trace is invariant. Since the real part of the trace has this important property, we shall define a notation for it:
\begin{equation}\label{ReTrace}
  \mathfrak{R}\textrm{tr}[T] = \frac{1}{2}\left(\, \textrm{Tr}[T] + \ovl{\textrm{Tr}[T]} \,\right) =
 \frac{1}{2}\left(\, T_\alpha^{\;\;\alpha} + \ovl{T_\alpha^{\;\;\alpha}} \,\right)\,.
\end{equation}
By the same reasoning one can prove that the real part of the trace of a tensor $T^{\dot{\alpha}}_{\;\;\dot{\beta}}$, i.e. a tensor on the representation $\bl{\bar{2}}^{-1}\otimes \bl{\bar{2}}$, is also invariant by the action of $GL(2;\mathbb{H})$ and, in particular, by the action of $SL(2;\mathbb{H})$.
We can sum up the latter conclusions by means of the following schematic relations:
\begin{equation}\label{16=15+1}
  \bl{2}\otimes \bl{\bar{2}} = \bl{15} \oplus \bl{1} \quad \text{ and } \quad   \bl{\bar{2}}^{-1}\otimes \bl{\bar{2}} = \widetilde{\bl{15}} \oplus \bl{1}\,,
\end{equation}
which display the fact that the 16-dimensional representations of $SL(2;\mathbb{H})$ on the left hand sides of the latter equations split as the direct sum of two irreducible representations, one of dimension $15$ and the other of dimension $1$ (the trivial representation, in which all elements of the group are represented by the number $1$). The elements that carry the representation $\bl{15}$ have the index structure $\check{T}_\alpha^{\;\;\beta}$ with $\mf{R}\text{tr}[\check{T}]=0$, whereas the objects that carry the representation $\widetilde{\bl{15}}$ have the index structure $\hat{T}^{\dot{\alpha}}_{\;\;\dot{\beta}}$ with $\mf{R}\text{tr}[\hat{T}]=0$. As we shall prove later, it turns out that the representations $\bl{15}$ and $\widetilde{\bl{15}}$ are equivalent to each other.

Concerning higher rank tensors in the quaternionic formalism, they do not carry a representation of the group in general. For instance, in general, one cannot associate a representation of the group to the tensors of the form $T_{\alpha\dot{\beta} \gamma\dot{\delta}}$ in the full quaternionic formalism. However, one could pose the following question: once the rank two tensors $T^{\mu\nu}$, which have two vector indices, carry a representation of $SO(5,1)$ and since a vector $V^\mu$ is represented in this quaternionic formalism by $V_{\alpha\dot{\beta}}$, how could it be possible to introduce the representation $T^{\mu\nu}$ in the quaternionic spinorial formalism? The naive answer would be that the objects $T_{\alpha\dot{\beta} \gamma\dot{\delta}}$ would do the job. However, as just argued, those do not even carry a representation of $SL(2;\mathbb{H})$.
 The correct answer is that it is possible as long as we introduce a frame of spinors and use real coefficients to span a general tensor. For instance, let $\{\Phi^{\mf{i}}\}$ be a frame of 8 spinors in the representation $\bl{2}$, with the index $\mf{i}$ labeling the eight elements of the frame. For instance, we could choose
$$ \Phi^{\mf{m}} = \left[
                     \begin{array}{c}
                       \bl{q}_{\mf{m}} \\
                       0 \\
                     \end{array}
                   \right] \quad \text{ and }\quad
\Phi^{\mf{m}+4} = \left[
                     \begin{array}{c}
                       0 \\
                       \bl{q}_{\mf{m}} \\
                     \end{array}
                   \right]\,,
 $$
with $\mf{m}\in{1,2,3,4}$. Then a tensor $T_{\alpha\beta}$ can be written as
$$ T  = c_{\mf{ij}} \, \Phi^{\mf{i}}\otimes \Phi^{\mf{j}}\,, $$
where $c_{\mf{ij}}$ are real coefficients. The action of the group in this tensor is then defined as
$$ T  \overset{ Q}{\longrightarrow}  T' =  c_{\mf{ij}} \, (Q \Phi^{\mf{i}})\otimes (Q\Phi^{\mf{j}}) \,.$$
An analogous definition of group action can be done for all other higher rank tensors. However, we shall not follow this path here since this approach is not covariant and it does not take advantage of the quaternionic algebra, on the contrary. Hence, in the present article we shall stick to the tensor representations displayed in Eq. (\ref{TensorRep}), which are the ones that take full advantage of the quaternionic spinorial formalism.




\section{Complex vectors in the quaternionic formalism}\label{Sec.CompVec}

In four dimensions, the spin group associated to Minkowski metric is $SL(2,\mathbb{C})$ and a real vector is represented by a hermitian matrix $V_{\alpha\dot{\beta}}$. It turns out that a general $2\times 2$ complex matrix that is not hermitian also represents vector, but a complex vector instead of a real one. Note that the number of degrees of freedom match. A $2\times 2$ hermitian matrix over the complex field have four real degrees of freedom, which is the number of degrees of freedom of a real vector in four dimensions, whereas a general complex matrix, not necessarily hermitian, has eight real degrees of freedom, which is the number of real degrees of freedom of a complex vector. Thus, in four dimensions any $2\times 2$ complex matrix can be associated to vector. The same reasoning does not apply in the six-dimensional spinorial formalism over the quaternions. This is simple to understand by counting the number of degrees of freedom of the vectors in six-dimensional Minkowski spacetime and comparing with the degrees of freedom of a quaternionic $2\times 2$ matrix, as summarized below:

\begin{eqnarray}
\left\{
  \begin{array}{ll}
   \text{General real vector}  \; &\Rightarrow\; 6 \text{ real degrees of freedom},\\
  \text{Hermitian element of } M(2;\mathbb{H}) \; &\Rightarrow\; 6 \text{ real degrees of freedom},\\
  \text{General complex vector} \; &\Rightarrow\; 12 \text{ real degrees of freedom}, \\
  \text{General element of } M(2;\mathbb{H})\; &\Rightarrow\; 16 \text{ real degrees of freedom}.
  \end{array}
\right.
\end{eqnarray}
So, due to the mismatch of the degrees of freedom of a complex vector in comparison with those of a general matrix $M_{\alpha \dot{\beta}}$, we conclude that a complex vector is not represented by a general $2\times 2$ quaternionic matrix, differently from what happens in four dimensions. A natural question then shows up: how can we represent complex vectors in this quaternionic spinorial formalism? The answer is than any vector should be represented by hermitian matrices, but in order to represent a complex vector we need to introduce an extra
imaginary unity, $\bl{i}$, that is independent from the quaternionic elements $\bl{I}$, $\bl{J}$, $\bl{K}$ and that commutes with them. For instance, $\bl{iK}=\bl{Ki}$. As an example of a complex vector, let us consider $V^\mu = \delta_1^\mu +\bl{i} \delta_4^\mu $ which is represented by the following matrix:
\begin{equation}\label{Vcomplex}
  V =  \sigma_1 + \bl{i}\, \sigma_4  = \left[
                                        \begin{array}{cc}
                                          0 & \bl{i}-\bl{I}\, \\
                                          \bl{i}+\bl{I} & 0 \\
                                        \end{array}
                                      \right] =
\left(
  \begin{array}{c}
    \frac{\bl{i}-\bl{I}}{\sqrt{2}} \\
    \frac{\bl{i}+\bl{I}}{\sqrt{2}} \\
  \end{array}
\right)\otimes \left( \small{\frac{1-\bl{i\,I}}{\sqrt{2}}}\,\, \small{ \frac{1+\bl{i\,I}}{\sqrt{2}} } \right)
\,.
\end{equation}
Note that the latter vector is the tensor product of two spinors, which should be expected from the fact that $V^\mu$ is a null vector, $V^\mu V_\mu = 0$.
Summing up, in order to represent complex vectors we need to complexify the quaternionic algebra, so that the underlining algebra passes from four-dimensional, with a basis $\{\bl{q}_{\mf{m}}\}$, to an eight-dimensional algebra with the basis $\{\bl{q}_{\mf{m}},\bl{i}\bl{q}_{\mf{m}}\}$.
In this scenario one needs to distinguish between complex conjugation, denoted by $\star$, and quaternionic conjugation, denoted by $\ovl{\ph{a}}$. For instance,
$$  \bl{i}^\star = - \bl{i}\;,\;\;\bl{I}^\star =  \bl{I}\;,\;\;(\bl{iI})^\star = - \bl{iI} \;,\;\;\ovl{\bl{i}} =  \bl{i} \;,\;\;\ovl{\bl{I}} =  -\bl{I} \,. $$
Therefore, the matrix given in Eq. (\ref{Vcomplex}) is hermitian because it is invariant by the action of the quaternionic hermitian conjugation, i.e. the composition of quaternionic conjugation followed by transposition. Hence we can write $V^\dagger = V$. Note that the element $\bl{i}$ is not affected by the quaternionic hermitian conjugation. The definition of $\dagger$ adopted here should not to be confused with the usual complex hermitian conjugation. In fact, note that the matrix (\ref{Vcomplex}) is not hermitian with regard to the imaginary unit $\bl{i}$.  Note that while quaternions form a division algebra, meaning that every nonzero element has an inverse, the same is not true for the complexified quaternionic algebra. For instance, the element $(1+\bl{i\,I})$ has no inverse.

Note that the complexification of the vector space allows one to tackle other signatures, since it is possible to make a one-to-one correspondence between signature and reality condition of the vector frame \cite{Trautman}. For instance, if a vector space is endowed with a metric $\bl{\eta}$ of Lorentzian signature then this means that it is possible to find a frame $\{ E_\mu\}$  such that $\bl{\eta}(E_\mu,E_\nu) = \eta_{\mu\nu}$, where $\eta_{\mu\nu}=\text{diag}(-,+,+,\cdots, +)$ is the Minkowski metric. In addition, the vectors of such frame are all real, $E_\mu^\star = E_\mu$, since up to this point we are assuming a real vector space. However, complexifying this space and defining $\varepsilon_0 = \bl{i} E_0$ and $\varepsilon_i =  E_i$ it follows that $\bl{\eta}(\varepsilon_\mu,\varepsilon_\nu) = \delta_{\mu\nu}$, meaning that the frame $\{ \varepsilon_\mu\}$ is Euclidean. But the Euclidean frame is not real, once $ \varepsilon_0^\star = -\varepsilon_0 $. We can say that the reality condition of an Euclidean frame is given by $ \varepsilon_0^\star = -\varepsilon_0 $ and $ \varepsilon_i^\star = \varepsilon_i $. In the same fashion, every other signature is uniquely associated to a reality condition. In particular, since we have started with a real space of Lorentzian signature, the reality condition for Lorentzian frames is that every vector is real. In this example we have adopted orthogonal frames, but we could also use frames with other types of inner products. For example, a six-dimensional null frame $\{e_i,\theta^i\}$, with $i,j\in\{1,2,3\}$, is a frame such that all vectors are null with the only non-vanishing inner products being $\bl{\eta}(e_i,\theta^j) = \frac{1}{2}\delta_i^j$. By means of a Lorentzian frame, which is real, we can construct the following null frame:
\begin{align*}
  e_1 =  \frac{1}{2} \left( E_5 + E_0 \right) \;,\;\;  e_2 &=  \frac{1}{2} \left( E_2 + \bl{i}E_4 \right) \;,\;\;
 e_3 =  \frac{1}{2} \left( E_3 + \bl{i}E_1 \right)  \;,\\
   \theta^1 =  \frac{1}{2} \left( E_5 - E_0 \right) \;,\;\;
 \theta^2 &=  \frac{1}{2} \left( E_2 - \bl{i}E_4 \right) \;,\;\;
 \theta^3 =  \frac{1}{2} \left( E_3 - \bl{i}E_1 \right)\;,
\end{align*}
whose reality condition is $e_1^\star = e_1$, $\theta^{1\,\star}=\theta^1$, $e_2^\star = \theta^2$ and $e_3^\star= \theta^3$. On the other hand, in the Euclidean signature the reality condition of a null frame is $e_i^\star = \theta^i$ \cite{Bat-Book}. It is in this sense that the complexification of the vector space allows us to handle any signature, since we can keep track of the signature by means of the reality condition of the adopted frame.

Since we have already established that a general matrix $M_{\alpha\dot{\beta}}$ does not represent a vector, unless it is hermitian, we should discuss the meaning of the remaining degrees of freedom that are not associated to the hermitian part. It turns out that a general matrix $M$ can always be written as the sum of a hermitian matrix and an anti-hermitian matrix:
\begin{equation}\label{AntiHermit}
  M = \frac{1}{2}\underbrace{\lef M + M^\dagger  \rig}_{\text{hermitian, }\bl{6} } \,+\, \frac{1}{2}
\underbrace{\lef M - M^\dagger  \rig}_{\text{anti-hermitian, } \bl{10} } \,,
\end{equation}
where the numbers $\bl{6}$ and $\bl{10}$ on the right hand side refer to the number of degrees of freedom in each of the parts. Now, note that just as the hermitian property of a matrix on the representation $\bl{2}\otimes \bar{\bl{2}}$ is kept invariant by the action of the group $SL(2;\mathbb{H})$, the same holds for an anti-hermitian matrix. More explicitly, if $A_{\alpha\dot{\beta}}$ is anti-hermitian, namely $A^\dagger = - A$, then after the action of
$Q\in SL(2;\mathbb{H})$ the transformed matrix will be $Q A Q^\dagger$, which is also anti-hermitian. This means that the 16-dimensional representation $\bl{2}\otimes \bar{\bl{2}}$ is not irreducible, rather it can be split as the direct sum of a 6-dimensional representation, formed by the hermitian matrices, plus a 10-dimensional representation, formed by the anti-hermitian matrices. Schematically, this is written as
$$ \bl{2}\otimes \bar{\bl{2}} \,=\, \bl{6} \,\oplus\, \bl{10}\,.  $$
The representations $\bl{6}$ and $\bl{10}$ cannot be further split, they are irreducible.
In the same fashion, the 16-dimensional representation $ \bar{\bl{2}}^{-1} \otimes \bl{2}^{-1}$ reduces into the direct sum of a hermitian part, that is 6-dimensional and is denoted by $\widetilde{\bl{6}}$, plus an anti-hermitian part, that will be denoted by $\bl{10}'$. Schematically, we have:
$$ \bar{\bl{2}}^{-1} \otimes \bl{2}^{-1} \,=\, \widetilde{\bl{6}} \,\oplus\, \bl{10}'\,.  $$
As we shall prove in the sequel, the representations $\bl{6}$ and $\widetilde{\bl{6}}$ are equivalent to each other, but $ \bl{10}$ is not equivalent to $ \bl{10}'$.

\section{Lie algebra of $SL(2;\mathbb{H})$ and its physical interpretation}\label{Sec.LieAlgebra}

We have already proved that $SL(2;\mathbb{H})$ is a double cover for the group $SO(5,1)$. In the present section we shall reinforce this fact by showing that they have the same Lie algebra. First we will obtain the Lie algebra of $SL(2;\mathbb{H})$ and then we will make connection with the Lie algebra of the Lorentz group $SO(5,1)$. The latter connection, in turn, will be used give a physical interpretation for the action of each element of $SL(2;\mathbb{H})$.

In order to know which elements $\mathcal{L}\in M(2;\mathbb{H})$ form the Lie algebra of $SL(2;\mathbb{H})$ we must assume that $Q = \mathbb{I}_2 + \kappa\,\mathcal{L}$, where $\kappa$ is an infinitesimal real parameter and then impose the constraint that defines the group $SL(2;\mathbb{H})$, namely $\Delta(Q)=1$. This constraint should be imposed only up to first order in $\kappa$. Assuming that $\mathcal{L}$ has the general form
\begin{equation*}
  \mathcal{L} = \left[
                  \begin{array}{cc}
                    \bl{a} & \bl{b} \\
                    \bl{c} & \bl{d} \\
                  \end{array}
                \right]\,,
\end{equation*}
where $\bl{a},\bl{b},\bl{c}$, and $\bl{d}$ are, in principle, general quaternions, we conclude that the Dieudonn\'{e} determinant of $Q = \mathbb{I}_2 + \kappa\,\mathcal{L}$ is, up to first order in $\kappa$, given by
\begin{align*}
  \Delta(Q) =& | (1 + \kappa \bl{a})(1+\kappa \bl{d}) - (1 + \kappa \bl{a})(\kappa \bl{c})(1-\kappa \bl{a})(\kappa\bl{b}) | \\
=& | 1 + \kappa(\bl{a} + \bl{d}) | + O(\kappa^2) = 1 + \kappa\,\mf{R}\text{tr}(\mathcal{L})  + O(\kappa^2)\,,
\end{align*}
where $\mf{R}\text{tr}(\mathcal{L})$ is the real part of the trace of $\mathcal{L}$. Thus, the condition $\Delta(Q)=1$ imposes that $\mf{R}\text{tr}(\mathcal{L})=0$, as it was argued in \cite{Sudbery84}. Hence $\bl{b}$ and $\bl{c}$ can be completely arbitrary quaternions, whereas $\bl{a}$ and $\bl{d}$ are constrained by the equation $(\bl{a}+\bl{d})+(\ovl{\bl{a}} + \ovl{\bl{d}})=0$. Therefore, a general element of $sl(2;\mathbb{H})$, the Lie algebra of  $SL(2;\mathbb{H})$, is given by a linear combination, with real coefficients, of the following matrices:
\begin{align}
 & \mathcal{J}_{\mf{mn}} = \left[
                            \begin{array}{cc}
                              \bl{q}^+_{\mf{mn}} & 0 \\
                              0 & \bl{q}^-_{\mf{mn}} \\
                            \end{array}
                          \right]\,,\quad
\mathcal{O}_{\mf{m}} = \left[
                            \begin{array}{cc}
                              0 & \bar{\bl{q}}_{\mf{m}} \\
                              0 & 0 \\
                            \end{array}
                          \right]\,,\quad  \nonumber\\
&\mathcal{A}_{\mf{m}} = \left[
                            \begin{array}{cc}
                              0 & 0 \\
                              \bl{q}_{\mf{m}} & 0 \\
                            \end{array}
                          \right]\,,\quad
\mathcal{S} = \frac{1}{2}\left[
                            \begin{array}{cc}
                              1 & 0 \\
                              0 & -1 \\
                            \end{array}
                          \right]\,, \label{LieAlgebra}
\end{align}
where $\boldsymbol{q}_{\mathfrak{m}} =  (\bl{I},\bl{J},\bl{K}, \boldsymbol{1})$ is the quaternionic basis whereas $\bl{q}^\pm_{\mf{mn}}$ are the quaternions defined by
$$ \bl{q}^+_{\mathfrak{m}\mathfrak{n}} = -\frac{1}{4}
\lef \bar{\bl{q}}_{\mathfrak{m}} \bl{q}_{\mathfrak{n}}  - \bar{\bl{q}}_{\mathfrak{n}}\bl{q}_{\mathfrak{m}}   \rig
\;,\quad  \bl{q}^-_{\mathfrak{m}\mathfrak{n}} = -\frac{1}{4}
\lef \bl{q}_{\mathfrak{m}} \bar{\bl{q}}_{\mathfrak{n}}  - \bl{q}_{\mathfrak{n}} \bar{\bl{q}}_{\mathfrak{m}}   \rig  \,.$$
By the very definitions of $\bl{q}^\pm_{\mf{mn}}$, it follows immediately that $\ovl{\bl{q}^\pm_{\mf{mn}}} = - \bl{q}^\pm_{\mf{mn}}$, i.e. these elements are all purely quaternionic. Note also that $\bl{q}^\pm_{\mf{mn}}$ are skew symmetric in their indices. Therefore, $\mathcal{J}_{\mf{mn}}$ comprise a total of 6 linearly independent matrices. A general element of the Lie algebra is then given by
\begin{equation}\label{GeneralL}
  \mathcal{L} = \lambda^{\mathfrak{m}\mathfrak{n}}\, \mathcal{J}_{\mathfrak{m}\mathfrak{n}} \,+\, b^{\mathfrak{m}} \,\mathcal{O}_{\mathfrak{m}}
\,+\, c^{\mathfrak{m}} \,\mathcal{A}_{\mathfrak{m}} \,+\, \omega\,\mathcal{S} \,,
\end{equation}
where the coefficients $\lambda^{\mathfrak{m}\mathfrak{n}} = -\lambda^{\mf{nm}}$, $b^{\mathfrak{m}}$, $c^{\mathfrak{m}}$, and $\omega$ are all real, summing a total of 15 degrees of freedom, which is the dimension of the groups $SL(2;\mathbb{H})$ and $SO(5,1)$. It is also possible to use this quaternionic spinorial formalism to represent the elements of $SO(6;\mathbb{C})$, the complexified version of the orthogonal group in six dimensions. In this case the coefficients  $\lambda^{\mathfrak{m}\mathfrak{n}}$, $b^{\mathfrak{m}}$, $c^{\mathfrak{m}}$, and $\omega$ can be complex but with respect a new imaginary unit $\bl{i}$, that is independent from $\bl{q}_{\mf{m}}$, as discussed when we digressed about complex vectors.

One can check that the matrices in Eq. (\ref{LieAlgebra}) obey the following algebra:
\begin{eqnarray}\label{SOA}
\left\{
  \begin{array}{ll}
    \left[\mathcal{J}_{\mathfrak{m}\mathfrak{n}}, \mathcal{J}_{\mathfrak{p}\mathfrak{q}} \right] = \delta_{\mathfrak{m}\mathfrak{p}}\mathcal{J}_{\mathfrak{n}\mathfrak{q}} + \delta_{\mathfrak{n}\mathfrak{q}}\mathcal{J}_{\mathfrak{m}\mathfrak{p}} - \delta_{\mathfrak{n}\mathfrak{p}}\mathcal{J}_{\mathfrak{m}\mathfrak{q}} - \delta_{\mathfrak{m}\mathfrak{q}}\mathcal{J}_{\mathfrak{n}\mathfrak{p}} \,, \\
\\
\left[\mathcal{J}_{\mathfrak{m}\mathfrak{n}}, \mathcal{O}_{\mathfrak{p}} \right] = \delta_{\mathfrak{m}\mathfrak{p}}\mathcal{O}_{\mathfrak{n}} - \delta_{\mathfrak{n}\mathfrak{p}}\mathcal{O}_{\mathfrak{m}}  \quad , \quad \left[\mathcal{J}_{\mathfrak{m}\mathfrak{n}}, \mathcal{A}_{\mathfrak{p}} \right] =  \delta_{\mathfrak{m}\mathfrak{p}}\mathcal{A}_{\mathfrak{n}} - \delta_{\mathfrak{n}\mathfrak{p}}\mathcal{A}_{\mathfrak{m}} \,, \\
\\
\frac{1}{2}\left[\mathcal{O}_{\mathfrak{m}}, \mathcal{A}_{\mathfrak{n}} \right] = \delta_{\mathfrak{m}\mathfrak{n}}\,\mathcal{S} - \mathcal{J}_{\mathfrak{m}\mathfrak{n}} \quad , \quad \left[\mathcal{O}_{\mathfrak{m}}, \mathcal{S} \right] = - \mathcal{O}_{\mathfrak{m}} \quad , \quad \left[\mathcal{A}_{\mathfrak{m}}, \mathcal{S} \right] = \mathcal{A}_{\mathfrak{m}} \,,  \\
\\
\left[\mathcal{O}_{\mathfrak{m}}, \mathcal{O}_{\mathfrak{n}} \right] = 0 \quad , \quad \left[\mathcal{A}_{\mathfrak{m}}, \mathcal{A}_{\mathfrak{n}} \right] = 0 \quad , \quad \left[\mathcal{J}_{\mathfrak{m}\mathfrak{n}}, \mathcal{S}\right] = 0 \,.
  \end{array}
\right.
\end{eqnarray}
Since $SL(2;\mathbb{H})$ is a double cover for the Lorentz group in six dimensions, both groups might lead to the same Lie algebra. Indeed, the Lie algebra of $SO(5,1)$ is generated by the bivectors $L_{\mu\nu} = - L_{\nu\mu}$ whose algebra is well known to be
\begin{equation}\label{AM}
\left[L_{\mu\nu}, L_{\rho\sigma} \right] = \eta_{\mu\rho}L_{\nu\sigma} + \eta_{\nu\sigma}L_{\mu\rho} - \eta_{\nu\rho}L_{\mu\sigma} - \eta_{\mu\sigma}L_{\nu\rho}  \,,
\end{equation}
where the Greek indices $\mu,\nu,\cdots$ run through the set $\{0,1,\cdots,5\}$. Thus, we can split these six-dimensional indices as $\mu \in \{0,\mathfrak{m},5\}$, where we shall recall that $\mathfrak{m},\mathfrak{n}\in\{1,2,3,4\}$. Doing so, we can find the following compatible identifications between the elements of the Lie algebras of $SO(5,1)$ and $SL(2;\mathbb{H})$:
\begin{equation}\label{idem}
\mathcal{J}_{\mathfrak{m}\mathfrak{n}} \leftrightarrow L_{\mathfrak{m}\mathfrak{n}} \quad , \quad \mathcal{O}_{\mathfrak{m}} \leftrightarrow  L_{0\mathfrak{m}} + L_{5\mathfrak{m}} \quad , \quad \mathcal{A}_{\mathfrak{m}} \leftrightarrow  L_{0\mathfrak{m}} - L_{5\mathfrak{m}}  \quad , \quad \mathcal{S} \leftrightarrow L_{50} \,.
\end{equation}
For instance, using these identifications along with the algebra (\ref{AM}) we can attain the  following result
\begin{align*}
   \left[\mathcal{J}_{\mathfrak{m}\mathfrak{n}}, \mathcal{J}_{\mathfrak{p}\mathfrak{q}} \right]  \leftrightarrow
\left[L_{\mathfrak{m}\mathfrak{n}}, L_{\mathfrak{p}\mathfrak{q}} \right] =&\;
 \delta_{\mathfrak{m}\mathfrak{p}}L_{\mathfrak{n}\mathfrak{q}} + \delta_{\mathfrak{n}\mathfrak{q}}L_{\mathfrak{m}\mathfrak{p}} - \delta_{\mathfrak{n}\mathfrak{p}} L_{\mathfrak{m}\mathfrak{q}} - \delta_{\mathfrak{m}\mathfrak{q}} L_{\mathfrak{n}\mathfrak{p}}\\
\leftrightarrow&\; \delta_{\mathfrak{m}\mathfrak{p}}\mathcal{J}_{\mathfrak{n}\mathfrak{q}} + \delta_{\mathfrak{n}\mathfrak{q}}\mathcal{J}_{\mathfrak{m}\mathfrak{p}} - \delta_{\mathfrak{n}\mathfrak{p}}\mathcal{J}_{\mathfrak{m}\mathfrak{q}} - \delta_{\mathfrak{m}\mathfrak{q}}\mathcal{J}_{\mathfrak{n}\mathfrak{p}}\,,
\end{align*}
with is in perfect accordance with the algebra (\ref{SOA}). Likewise, note that
$$ \frac{1}{2} \left[\mathcal{O}_{\mathfrak{m}}, \mathcal{A}_{\mathfrak{n}}\right] \leftrightarrow \frac{1}{2} \left[L_{0\mathfrak{m}} + L_{5\mathfrak{m}}, L_{0\mathfrak{n}} - L_{5\mathfrak{n}} \right]
=  \delta_{\mathfrak{m}\mathfrak{n}}\, L_{50}  -  L_{\mathfrak{m}\mathfrak{n}} \leftrightarrow
\delta_{\mathfrak{m}\mathfrak{n}}\, \mathcal{S}  -  \mathcal{J}_{\mathfrak{m}\mathfrak{n}} \,,$$
which, again, is in perfect accordance with Eq. (\ref{SOA}). The remaining commutation relations in Eq. (\ref{SOA}) can be retrieved in the same fashion.

The Lorentz transformations generated by $L_{\mu\nu}$ have simple interpretations. Indeed, if $\{E_\mu\}$ is the adopted Lorentz frame, so that
$\bl{\eta}(E_\mu , E_\nu)=\eta_{\mu\nu}$, with $\bl{\eta}$ standing for the metric tensor in $\mathbb{R}^{5,1}$, then it follows that $L_{0i}$ generates Lorentz boosts along the spatial direction $E_i$ whereas $L_{ij}$ is the generator of rotations in the plane spanned by the space-like vectors $E_i$ and $E_j$.  Using the identifications (\ref{idem}) we can then give physical meanings for the transformations generated by $\mathcal{J}_{\mathfrak{m}\mathfrak{n}}$, $\mathcal{O}_{\mathfrak{m}}$, $\mathcal{A}_{\mathfrak{m}}$ and $\mathcal{S}$. For instance, due to the relation $\mathcal{J}_{\mathfrak{m}\mathfrak{n}} \leftrightarrow L_{\mathfrak{m}\mathfrak{n}}$ it follows that $\mathcal{J}_{\mathfrak{m}\mathfrak{n}}$ should be interpret as the generator of the rotations on the plane $E_{\mathfrak{m}}\wedge E_{\mathfrak{n}}$. Analogously, $\mathcal{S}$ generates a boost in the direction $E_5$. Regarding $\mathcal{O}_{\mathfrak{m}}$ and  $\mathcal{A}_{\mathfrak{m}}$ they are a composition of a boost and a rotation. More precisely, the latter transformations are known as null rotations, which are rotations that keep specific null directions invariant. For instance, let us consider  $\mathcal{O}_{\mf{m}}$. The action of $L_{\mu\nu}$ on the position vector $x^\mu$ can be obtained if we adopt the identification $L_{\mu\nu}\sim x_\mu \partial_\nu - x_\nu \partial_\mu$. Thus, the action of $\mathcal{O}_{\mathfrak{m}}$ in the latter position vector is given by
\begin{align}
  \mathcal{O}_{\mathfrak{m}} x^\mu & \leftrightarrow (L_{0\mathfrak{m}} + L_{5\mathfrak{m}} ) x^\mu \nonumber \\
& \sim (x_0 \partial_{\mathfrak{m}}- x_{\mathfrak{m}} \partial_0 + x_5 \partial_{\mathfrak{m}} - x_{\mathfrak{m}} \partial_5  )x^\mu  =
(x^5-x^0) \delta_{\mathfrak{m}}^\mu - x^{\mathfrak{m}} (\delta_0^\mu +  \delta_5^\mu) . \label{OmAction}
\end{align}
So, if we consider the position vectors along the null direction $e_1 = \frac{1}{2}(E_5 +E_0 )$, namely those such that $x^{\mathfrak{m}}=0$ and $x^0=x^5$ we have that the right hand side of Eq. (\ref{OmAction}) vanishes, meaning that these directions are invariant by the action of $\mathcal{O}_{\mathfrak{m}}$. Likewise, one can prove that the action of $\mathcal{A}_{\mathfrak{m}}$ keeps the null direction
$\theta^1 = \frac{1}{2}( E_5 - E_0 )$ invariant. Thus, the elements $\text{exp}\left[c^{\mathfrak{m}}\mathcal{A}_{\mathfrak{m}}\right]\in SL(2;\mathbb{H})$ provide all transformation that keep the vector $\theta^1$ invariant.

There is another way to check the physical interpretation of the transformations of $SL(2; \mathbb{H})$, namely by directly computing their action on the vectors through the spinorial formalism. We know that a vector $V^\mu$ can be represented by a $2\times 2$ quaternionic matrix $V$ that is hermitian, $V^\dagger = V$, i.e.
\begin{equation}\label{HM}
V = \begin{bmatrix}
\ell & \overline{\boldsymbol{v}}\\
\boldsymbol{v} & n
\end{bmatrix} \,,
\end{equation}
where $\ell$ and $n$ are real while $\bl{v}$ can be a full quaternion. More precisely, these components are related to $V^\mu$ by the following relations:
\begin{equation}\label{RelationellV}
   \ell = (V^0 + V^5) \;,\quad  n = (V^0 - V^5) \;,\quad  \bl{v} =  V^1 \bl{I} + V^2 \bl{J}+ V^3 \bl{K} + V^4 = V^{\mathfrak{m}}\,q_{\mathfrak{m}} \,.
\end{equation}
 The action of an element $Q\in SL(2; \mathbb{H})$ on the vector is then given by $V \rightarrow Q V Q^{\dagger}$. Thus, let us compute the finite transformations associated to the generators $\mathcal{S}$, $\mathcal{O}_{\mathfrak{m}}$, $\mathcal{A}_{\mathfrak{m}}$, and $\mathcal{J}_{\mathfrak{m}\mathfrak{n}}$ and then check how these transform a general vector $V$. Let us start with $\mathcal{S}$, whose exponential is given by
$$ Q_{\omega} = e^{\omega\,\mathcal{S}} = \begin{bmatrix}
e^{\omega/2} & 0\\
0 & e^{-\omega /2}
\end{bmatrix} \,.$$
Its action on $V$ is then given by
$$ V \rightarrow V' = Q_{\omega} V  Q_{\omega}^\dagger =
\begin{bmatrix}
\ell e^\omega & \overline{\boldsymbol{v}}\\
\boldsymbol{v} & n e^{-\omega}
\end{bmatrix}  \;
\left\{
  \begin{array}{ll}
    \ell \rightarrow \ell' = e^\omega\,\ell \\
    n \rightarrow n' = e^{-\omega}\,n \\
    \boldsymbol{v} \rightarrow \bl{v}'= \bl{v}
  \end{array}
\right.\,.
 $$
Then, using the relations (\ref{RelationellV}) we find that the components of the vector after the transformation are given by:
$$ V^{0'}  = \cosh\omega V^0 + \sinh\omega V^5\;,\quad
V^{\mathfrak{m}'}  =  V^{\mathfrak{m}}\;,\quad
 V^{5'}  = \sinh\omega V^0 + \cosh\omega V^5\,.  $$
This is clearly a Lorentz boost along the direction $E_5$, with $\omega$ being the so called rapidity parameter. The speed of the boost is given by $|\tanh\omega|$.

Consider now the Lie algebra element $b^{\mathfrak{m}}\mathcal{O}_{\mathfrak{m}}$. Its exponential is given by
\begin{equation*}
  Q_{\bl{b}} = e^{b^{\mathfrak{m}}\mathcal{O}_{\mathfrak{m}}} =
\begin{bmatrix}
1 & \bar{\boldsymbol{b}}\,\\
0 & 1\,
\end{bmatrix}\,,
\end{equation*}
where $\bl{b} = b^{\mathfrak{m}} \bl{q}_{\mf{m}}$, which lead to the following transformations on the components of the vector
\begin{equation}
\ell'= \ell + \overline{\boldsymbol{b}} \boldsymbol{v}+ \overline{\boldsymbol{v}} \boldsymbol{b} +  |\boldsymbol{b}|^{2}\,n\;,\quad
n' =n \;,\quad
\boldsymbol{v}' = \boldsymbol{v} + \boldsymbol{b}\,n \,.
\end{equation}
Thus, vectors such $n=0$ and $\bl{v}=0$ are kept invariant by the action of $Q_{\bl{b}}$. These vectors are the ones such that $V^0=V^5$ and $V^{\mathfrak{m}}=0$, i.e. they point in the null direction $e_1 = \frac{1}{2}( E_5+E_0 )$. This interpretation for the Lorentz transformations generated by $\mathcal{O}_{\mathfrak{m}}$ is in perfect accordance with what was obtained previously by means of identifications of the Lie algebras of $SL(2;\mathbb{H})$ and $SO(5,1)$. Likewise, one can check that the finite transformation associated to $c^{\mf{m}}\mathcal{A}_{\mf{m}}$, namely $Q_{\bl{c}}=e^{c^{\mf{m}}\mathcal{A}_{\mf{m}}}$, yields the following transformation on the vector components:
\begin{equation}
\ell'= \ell  \;,\quad  n' = n + \boldsymbol{v} \overline{\boldsymbol{c}} + \boldsymbol{c} \overline{\boldsymbol{v}} +  |\boldsymbol{c}|^{2}\ell
\;,\quad \bl{v}' =\boldsymbol{v} + \boldsymbol{c}\,\ell \,,
\end{equation}
where $\bl{c}=c^{\mf{m}}\bl{q}_{\mf{m}}$. Thus, vectors such that $\ell=0$ and $\bl{v}=0$ are invariant by the latter transformation, meaning that the null vector $\theta^1 =\frac{1}{2} ( E_5- E_0)$ is unchanged by the action of  $Q_{\bl{c}}$. In the jargon, it is said that $Q_{\bl{b}}$ implements a null rotation around $e_1$, while $Q_{\bl{c}}$ yields a null rotation around $\theta^1$.

Finally, it remains to analyse the effect of the transformations generated by $\mathcal{J}_{\mf{mn}}$. Defining its associated finite transformation by
$Q_{\lambda} = \text{exp}[\lambda^{\mf{mn}} \mathcal{J}_{\mf{mn}}]$, it follows that the transformation of the vector components are given by
$$ \ell'=\ell \;,\quad  n' = n \;,\quad  \bl{v}' = e^{\lambda^{\mf{mn}}\bl{q}^-_{\mf{mn}}}\,\bl{v}\,e^{-\lambda^{\mf{mn}}\bl{q}^+_{\mf{mn}}} \,. $$
Thus, the vectors $E_0$ and $E_5$ are invariant by the latter transformation, while the four spacelike vectors $E_{\mathfrak{m}}$ can be changed. This means that $\mathcal{J}_{\mf{mn}}$ generate rotations on the four-dimensional space generated by $\{E_{\mathfrak{m}}\}$. For instance, let us consider the action of the transformation generated by $\mathcal{J}_{12}$. Noting that $(\bl{q}^+_{12},\bl{q}^-_{12} )  = (\frac{1}{2}\bl{K},\frac{1}{2}\bl{K})$, it follows that the action of $e^{\lambda \mathcal{J}_{12}}$ is such that
$$ \bl{v}' = e^{\lambda\bl{K}/2}\,\bl{v}\,e^{-\lambda\bl{K}/2} =
 V^1 e^{\lambda\bl{K}/2}\,\bl{I}\,e^{-\lambda\bl{K}/2}+ V^2 e^{\lambda\bl{K}/2}\,\bl{J}\,e^{-\lambda\bl{K}/2}
+ V^3\bl{K} + V^4 \,,$$
where it has been used the fact that $e^{\lambda\bl{K}/2}$ commutes with $\bl{K}$ and $1$, whereas $\bl{I}$ and $\bl{J}$ cannot pass through this exponential. In fact, we have that
\begin{align*}
  e^{\lambda\bl{K}/2}\,\bl{I}\,e^{-\lambda\bl{K}/2} & = \left[ \cos(\lambda/2) + \bl{K} \sin(\lambda/2) \right]\bl{I}
\left[ \cos(\lambda/2) - \bl{K} \sin(\lambda/2) \right]\\
& = \cos(\lambda)\,\bl{I} + \sin(\lambda)\bl{J}\,,
\end{align*}
with an analogous relation holding for the action of the exponentials on $\bl{J}$. At the end of the day, the following transformation rules are obtained
$$ V^{1'} = \cos(\lambda) V^1 - \sin(\lambda)V^2\,,\;  V^{2'} = \cos(\lambda) V^2 + \sin(\lambda)V^1\,,\;
V^{3'} =  V^3 \,,\;V^{4'} = V^4\,.   $$
This proves that  $\mathcal{J}_{12}$ does, indeed, generate rotations on the plane $E_1\wedge E_2$.

\section{The map $V\rightarrow \widetilde{V}$}\label{Sec.MapTilde}

Since a vector admits two kinds of representations in the $SL(2;\mathbb{H})$ spinorial formalism, there might exist a canonical map connecting $V_{\alpha\dot{\beta}}$ to $\widetilde{V}^{\dot{\alpha}\beta}$. Indeed, using Eq. (\ref{VVtil}), it follows that this map is given by
\begin{equation}\label{VVtil2}
V  \overset{ \sim }{\longrightarrow} \widetilde{V} =  -\textrm{det}(V)\,V^{-1} \quad
\text{and} \quad  \widetilde{V}   \overset{ \sim }{\longrightarrow} \widetilde{\widetilde{V}} =  -\textrm{det}(\widetilde{V})\,\widetilde{V}^{-1} = V\,.
\end{equation}
In terms of indices, we have
$$ V_{\alpha\dot{\beta}}  \overset{ \sim }{\longrightarrow} \widetilde{V}^{\dot{\alpha}\beta}
\quad
\text{and} \quad   \widetilde{V}^{\dot{\alpha}\beta} \overset{ \sim }{\longrightarrow} \widetilde{\widetilde{V}}_{\alpha\dot{\beta}}=V_{\alpha\dot{\beta}}  \,. $$
Note that the operation $\sim$ maps hermitian matrices into hermitian matrices. One interesting feature of this map is that, in spite of the fact that the determinant and the inverse of a matrix are non-linear operations on the matrix, it turns out that the final action of the map $\sim$ is linear in the components of the vector. Namely
\begin{equation}\label{VVtil3}
   V = V^\mu \sigma_\mu  \;\;  \Leftrightarrow\;\;  \widetilde{V} = V^\mu \,\widetilde{\sigma}_\mu \,.
\end{equation}
In particular, using the latter relation one can easily obtain the tilded version of any vector, without the trouble of having to compute the inverse and the determinant of the matrix, as it would be required if we used the initial definition given in Eq. (\ref{VVtil2}). Moreover, as given in Eq. (\ref{VVtil2}), the map $\sim$ is ill-defined for light-like vectors, since in such case the determinant vanishes and the matrix is not invertible. On the other hand, if we use Eq. (\ref{VVtil3}) the fact that a vector is light-like is not an issue whatsoever. For instance, consider the light-like vector given by
$$ V = \mathrm{v}\cosh\theta\,\sigma_0 + \mathrm{v}\sinh\theta\,\sigma_5 + v_1\, \sigma_1 + v_2 \,\sigma_2 + v_3 \,\sigma_3 +v_4 \,\sigma_4\;,$$
where $\mathrm{v} \equiv \sqrt{v_1^2 + v_2^2 + v_3^2 + v_4^2}$ and $\theta \in \mathbb{R}$. Then, defining the quaternion
$$\bl{v} = v_{\mf{m}} \bl{q}_{\mf{m}} =  v_1 \bl{I} + v_2\bl{J} + v_3 \bl{K} + v_4 \,,$$
and using Eq. (\ref{VVtil3}), it follows that the map $\sim$ yields:
\begin{equation*}
  V = \left[
          \begin{array}{cc}
            \mathrm{v}e^\theta & \bar{\bl{v}}  \\
            \bl{v} & \mathrm{v} e^{-\theta}\\
          \end{array}
        \right]
         \; \overset{ \sim }{\longrightarrow} \;\\
           \widetilde{V}  =  \left[
          \begin{array}{cc}
            -\mathrm{v}e^{-\theta} & \bar{\bl{v}} \\
            \bl{v} & -\mathrm{v} e^{\theta}\\
          \end{array}
        \right]   \,.
\end{equation*}
Since these matrices have vanishing determinants, they can be split as the direct product of a column-vector and a row-vector, as done below:
\begin{equation*}
  V =  \left[
                    \begin{array}{c}
                      1 \\
                      \frac{\bar{\bl{v}}}{\mathrm{v}}e^{-\theta} \\
                    \end{array}
                  \right]\otimes [\mathrm{v}e^\theta\;\;\bl{v}]
         \; \overset{ \sim }{\longrightarrow} \;\\
           \widetilde{V}  = \left[
                    \begin{array}{c}
                      -\bl{v} \\
                      \mathrm{v}e^\theta \\
                    \end{array}
                  \right]\otimes \Big[\frac{\bar{\bl{v}}}{\mathrm{v}}e^{-\theta}\;\;-1\Big]  \,.
\end{equation*}
Thus, defining the spinors
$$ \xi_\alpha =  \left[
                    \begin{array}{c}
                      1 \\
                      \frac{\bar{\bl{v}}}{\mathrm{v}}e^{-\theta} \\
                    \end{array}
                  \right] \quad \text{and} \quad
\chi_{\dot{\beta}}  = [\mathrm{v}e^\theta\;\;\bl{v}]\,,$$
it follows that the latter relation can be more compactly written as:
\begin{equation}\label{VVtil4}
  V_{\alpha\dot{\beta}} = \xi_\alpha\,\chi_{\dot{\beta}} \;\; \overset{ \sim }{\longrightarrow}  \;\; \widetilde{V}^{\dot{\alpha} \beta} = \tilde{\chi}^{\dot{\alpha}}\,\tilde{\xi}^{\beta}\,,
\end{equation}
where $\tilde{\chi}^{\dot{\alpha}}$ and $\tilde{\xi}^{\alpha}$ are given by
$$ \tilde{\chi}^{\dot{\alpha}} = \chi_{\dot{\beta}}\,\varepsilon^{\dot{\beta}\dot{\alpha}}\, \;\; \text{and}\;\;
 \tilde{\xi}^{\alpha} = \varepsilon^{\alpha\beta}\,\xi_{\beta}\;,\;\; \text{where}\;\;
 \varepsilon^{\dot{\alpha}\dot{\beta}} = \left[
                                           \begin{array}{cc}
                                             0 & 1 \\
                                             -1 & 0 \\
                                           \end{array}
                                         \right] = \varepsilon^{\alpha\beta}\,.
    $$
Actually, the relation (\ref{VVtil4}) can be generalized for the cases in which the vector is non-null. Indeed, using the fact that vectors are represented by hermitian matrices, we can prove by direct calculation that the following relations hold for an arbitrary vector:
\begin{equation}\label{VVtil5}
  \widetilde{V} = \varepsilon\, V^t\, \varepsilon = \varepsilon\, \ovl{V}\, \varepsilon \,.
\end{equation}
This linear relation proves that the representations $\bl{6}$ and $\widetilde{\bl{6}}$, which are the hermitian parts of the representations $\bl{2}\otimes \bar{\bl{2}}$ and  $\bar{\bl{2}}^{-1} \otimes \bl{2}^{-1}$, are equivalent.

It is worth keeping in mind that, in spite of the validity of Eq. (\ref{VVtil4}), the object $\varepsilon^{\alpha\beta}$ does not provide a map between the representations $\bl{2}$ and $\bl{2}^{-1}$. Likewise, $\varepsilon^{\dot{\alpha}\dot{\beta}}$ does not provide a map between $\bar{\bl{2}}$ and $\bar{\bl{2}}^{-1}$. Indeed, suppose that for every spinor $\xi_\alpha$ that transforms according to $\bl{2}$ we have $\tilde{\xi}^{\alpha}$ carrying the representation $\bar{\bl{2}}^{-1}$, so that
\begin{equation}\label{xitransf}
   \xi_\alpha \overset{ Q }{\longrightarrow} \xi'_\alpha = Q_\alpha^{\;\;\beta}\xi_\beta \;\; \text{ and }\;\;
\tilde{\xi}^\alpha \overset{ Q }{\longrightarrow} \tilde{\xi}'^{\alpha} = \tilde{\xi}^\beta (Q^{-1})_\beta^{\;\;\alpha}\,.
\end{equation}
Then, inserting the relations $\tilde{\xi}^{\alpha} = \varepsilon^{\alpha\beta}\xi_{\beta}$ and $\tilde{\xi}'^{\alpha} = \varepsilon^{\alpha\beta}\xi'_{\beta}$ into the definition of $\tilde{\xi}'^{\alpha}$, given in Eq. (\ref{xitransf}), we are led to
$$ \varepsilon^{\alpha\beta} \left( Q_\beta^{\;\;\sigma}\xi_\sigma \right) = \varepsilon^{\beta\sigma}\xi_{\sigma} (Q^{-1})_\beta^{\;\;\alpha}
\;\; \Rightarrow\;\; \varepsilon\,Q = (Q^{-1})^t\,\varepsilon \;\; \Rightarrow\;\;(Q^{-1})^t =  \varepsilon\,Q\,\varepsilon^{-1}\,, $$
where in the first implication sign we have assumed the particular case in which $\xi_\sigma$ is real, so that it commutes with $Q$.
However, the latter relation is not true for an arbitrary $Q\in SL(2;\mathbb{H})$. For example, if we take $Q=\textrm{diag}(\cos\theta + \bl{I} \sin\theta, \cos\varphi + \bl{J} \sin\varphi)$ we can check that $(Q^{-1})^t \neq  \varepsilon\,Q\,\varepsilon^{-1}$. This proves that although the map
$$  \xi_\alpha\,\chi_{\dot{\beta}} \rightarrow \tilde{\chi}^{\dot{\alpha}}\,\tilde{\xi}^{\beta}  $$
does, indeed, connect an object on the representation  $\bl{2}\otimes \bar{\bl{2}}$ to an object on the representation  $\bar{\bl{2}}^{-1} \otimes \bl{2}^{-1}$, it is not true that the map $ \xi_\alpha \rightarrow \tilde{\xi}^{\alpha}$ connects the representations $\bl{2}$ and $\bl{2}^{-1}$.

Up to now, we have digressed about the map $A\rightarrow \widetilde{A}$ just in the case in which the matrix $A$ is hermitian. What about non-hermitian matrices? In other words, is this map defined for general tensors in the representations $\bl{2}\otimes \bar{\bl{2}}$ and  $\bar{\bl{2}}^{-1} \otimes \bl{2}^{-1}$? The answer can be yes or no, depending on what is meant by such a map. First of all, based on the hermitian case, more precisely due to Eqs. (\ref{VVtil2}) and (\ref{VVtil5}), we could generalize the map $\sim$ to act in a general matrix $A$ in the following ways:
\begin{equation}\label{Atil}
  \left\{
    \begin{array}{ll}
      (i):\, &\widetilde{A} = -\Delta(A)A^{-1} \,,   \\
      (ii):\, &\widetilde{A} = \varepsilon\,A^t\,\varepsilon \,, \\
      (iii):\, &\widetilde{A} = \varepsilon\,\ovl{A}\,\varepsilon \,.
    \end{array}
  \right.
\end{equation}
Note that in the first map we have wrote $\Delta(A)$ instead of $\text{det}(A)$, differently from Eq. (\ref{VVtil2}). The reason for this change is that, as previously argued, the operation $\text{det}(A)$ is not well defined for a general matrix $A$, rather $\Delta(A)$ is the unique reasonable generalization of determinant for quaternionic matrices. When $A$ is hermitian, the three definitions in Eq. (\ref{Atil}) coincide. However, for a general case they all differ. Let us start analysing the option $(i)$. As defined in $(i)$, the map $\sim$ takes the representation $\bl{2}\otimes \bar{\bl{2}}$ into  $\bar{\bl{2}}^{-1} \otimes \bl{2}^{-1}$ and vice versa, which is a desirable property. However, in spite of the fact that $\widetilde{(\lambda A)} = \lambda \widetilde{A}$ for any $\lambda\in \mathbb{R}$, this map is not linear on the components of $A$, rather it is a fractional relation that is homogeneous of degree 1 on the entries of $A$. Therefore, from the group theory point of view, such a map does not provide an equivalence relation between the representations  $\bl{2}\otimes \bar{\bl{2}}$ and  $\bar{\bl{2}}^{-1} \otimes \bl{2}^{-1}$ of $SL(2;\mathbb{H})$. On the other hand, whereas the options $(ii)$ and $(iii)$ for the map $\sim$ are clearly linear functions of the entries of $A$, they do not map elements in the representation $\bl{2}\otimes \bar{\bl{2}}$ into objects that transform according to $\bar{\bl{2}}^{-1} \otimes \bl{2}^{-1}$, and vice versa. In order to understand this assertion, let us suppose that $\widetilde{A} = \varepsilon A^\diamond \varepsilon$, where the operation $\diamond$ is either transposition or quaternionic conjugation. Then, if $A$ carries the representation $\bl{2}\otimes \bar{\bl{2}}$ of $SL(2;\mathbb{H})$, it follows $\widetilde{A}$ will transform as
$$ A \overset{ Q }{\longrightarrow} \widetilde{A}' = \varepsilon \,(A' )^\diamond\,\varepsilon = \varepsilon \,(Q\,A\,Q^{\dagger} )^\diamond\,\varepsilon  \,.$$
On the other hand, if $\widetilde{A}$ is on the representation $\bar{\bl{2}}^{-1} \otimes \bl{2}^{-1}$ then it should transform as
$$ A \overset{ Q }{\longrightarrow} \widetilde{A}'  = Q^{\dagger-1}\widetilde{A}  \,Q^{-1}
 = Q^{\dagger-1}\varepsilon \,A^\diamond\,\varepsilon \,Q^{-1} \,. $$
Thus, equating both expressions for $\widetilde{A}'$ we end up with the following identity
\begin{equation}\label{IdentityA}
   \varepsilon \,(Q\,A\,Q^{\dagger} )^\diamond\,\varepsilon = Q^{\dagger-1}\varepsilon \,A^\diamond\,\varepsilon \,Q^{-1} \,,
\end{equation}
which should be valid for an arbitrary $A\in M(2;\mathbb{H})$ and an arbitrary $Q\in SL(2;\mathbb{H})$. As a sidenote, note that on the left hand side of the above equation we cannot replace
$(Q A Q^{\dagger} )^\diamond$ by $Q^{\dagger\diamond} A^\diamond Q^\diamond$, since this property holds neither for transposition nor for quaternionic conjugation when the matrices have quaternionic entries. However, as a matter of fact, the identity (\ref{IdentityA}) is not true. For example, assuming
$$  A = \left[
          \begin{array}{cc}
            0 & \bl{I} \\
            0 & 0 \\
          \end{array}
        \right] \quad \text{ and } \quad
 Q = \left[
          \begin{array}{cc}
            1 & \bl{J} \\
            0 & 1 \\
          \end{array}
        \right]
 $$
one can check that the equality (\ref{IdentityA}) does not hold. Nonetheless, in the special case in which $A$ is hermitian, it can be proved that Eq. (\ref{IdentityA}) is valid for any choice of $Q\in SL(2;\mathbb{H})$. The proof of the latter assertion is somehow involved and the easier path to attain it is by assuming that $Q = \mathbb{I}_2 + \kappa\,\mathcal{L}$, where $\kappa$ is an infinitesimal real parameter and $\mathcal{L}$ is an arbitrary element of the Lie algebra of $SL(2;\mathbb{H})$, namely $\mathcal{L}$ is an arbitrary element of $M(2;\mathbb{H})$ such that $\mf{R}\textrm{tr}[\mathcal{L}]=0$. Summing up, we conclude that none of the definitions $(ii)$ and $(iii)$ for the map $\sim$ yield a map that connects the representations $\bl{2}\otimes \bar{\bl{2}}$ and $\bar{\bl{2}}^{-1} \otimes \bl{2}^{-1}$. Thus, the map $\sim$ lead to an equivalence between the these two representations only when restricted to hermitian matrices, but not otherwise. In particular, the anti-hermitian representations $\bl{10}$ and $\bl{10}'$, defined in Sec. \ref{Sec.CompVec}, are not equivalent.

\section{Bivectors and 3-vectors}\label{Sec.Bivec3Vec}

Now that we know how to represent vectors in the quaternionic spinorial formalism, we should move on and try to obtain the spinorial representations carried by arbitrary tensors. Since any tensor $T^{\mu_1\cdots \mu_n}$ can be spanned by the tensor product of vectors, the task should be simple. Nevertheless, as previously explained, most tensors cannot be represented in a covariant way in this quaternionic spinorial formalism. Indeed, the only tensor representations that are covariant and take full advantage of the quaternionic formalism are the ones displayed in Eq. (\ref{TensorRep}). In this section we shall prove that these representations comprise exactly the vectors, bivectors, and 3-vectors. More explicitly, these representations decompose in the following irreducible parts:
\begin{equation}\label{TenRepIrred}
  \left\{
  \begin{array}{ll}
    T_{\alpha\dot{\beta}}:& \bl{2}\otimes\bar{\bl{2}} = \bl{6}\oplus \bl{10} \,, \\
    T_{\alpha}^{\;\;{\beta}}:& \bl{2}\otimes \bl{2}^{-1} = \bl{15}\oplus \bl{1} \,,\\
T^{\dot{\alpha}\beta}:& \bar{\bl{2}}^{-1}  \otimes \bl{2}^{-1} = \widetilde{\bl{6}}\oplus \bl{10}' \,,\\
T^{\dot{\alpha}}_{\;\;\,\dot{\beta}}:& \bar{\bl{2}}^{-1}\otimes \bar{\bl{2}}= \widetilde{\bl{15}}\oplus \bl{1}\,.
  \end{array}
\right.
\end{equation}
As already proved, the representations $\bl{6}$ and $\widetilde{\bl{6}}$ are equivalent to each other and they represent the vectors in the spinorial formalism. In this section we shall see that $\bl{15}$ and $\widetilde{\bl{15}}$ are also equivalent to each other and they represent bivectors, i.e. skew symmetric tensors of rank two ($B^{\mu\nu}=B^{[\mu\nu]}$). In turn, the 3-vectors, which are totally skew symmetric tensors of rank three ($T^{\mu\nu\lambda} = T^{[\mu\nu\lambda]}$), carry the representation $\bl{10}\oplus\bl{10}'$. Once proved the latter assertions, we have accomplished the task of giving a meaning for all irreducible representations displayed on the right hand side of Eq. (\ref{TenRepIrred}). We should not bother about the representation $\bl{1}$, which is just the trivial representation, i.e. the objects carrying such representation are invariant under the action of the group, also known as scalars.

In six dimensions the number degrees of freedom of a bivector, namely a skew symmetric tensor of rank two $B_{\mu\nu}=-B_{\nu\mu}$, is 15. This is exactly the dimension of the representations  $\bl{15}$ and $\widetilde{\bl{15}}$. Moreover, in any dimension different from 4, the bivectors carry a representation of the Lorentz group that is irreducible.  Therefore, it is reasonable to guess that the spinorial representation of the bivectors of $SO(5,1)$ is given by  $\bl{15}$ or by $\widetilde{\bl{15}}$. Indeed, there are several ways to arrive at this conclusion. For instance, since the generators of the orthogonal groups, like $SO(5,1)$, are the bivectors, it follows that the generators of the spin group must carry the same spinorial representation of the bivectors. As previously proved, the generators of $SL(2;\mathbb{H})$ are matrices whose real part of the trace must vanish, in accordance with the definition of the representations $\bl{15}$ and $\widetilde{\bl{15}}$. Note also that the elements of the group, and, therefore, the elements of the Lie algebra as well, when acting on a spinor must yield a spinor of the same type, by the very definition of the spinorial representations. Thus, when acting on a spinor $\psi_\alpha$, that carries the representation $\bl{2}$, a bivector must yield a spinor on the same representation. Taking into account that the action of the group on this spinorial representation is on the left, we conclude that the bivector must have the index structure $B_\alpha^{\;\;\beta}$, so that when acting on $\psi_\alpha$ by the left we have $B_\alpha^{\;\;\beta}\psi_\beta$, which is an object on the representation $\bl{2}$.  Likewise, the same bivector when acting on a spinor $\phi^\alpha$, that carries the representation $\bl{2}^{-1}$, yields $\phi^\beta B_\beta^{\;\;\alpha}$, since for this representation the action is on the right. Thus, we conclude that the bivectors are represented in the spinorial formalism by objects with the index structure $B_\alpha^{\;\;\beta}$ and such that $\mf{R}\text{tr}(B)=0$. This proves that bivectors carry the representation $\bl{15}$. However, we could also consider the action of the Lie algebra, i.e. the action of the bivectors, on the spinors $\xi_{\dot{\alpha}}$ and $\varphi^{\dot{\alpha}}$, carrying the representations $\bar{\bl{2}}$ and $\bar{\bl{2}}^{-1}$. In the latter case, in order to preserve the spinor type, the element of the Lie algebra must have the index structure $B^{\dot{\alpha}}_{\;\;\dot{\beta}}$, namely it must be on the representation $\widetilde{\bl{15}}$. Thus, we conclude that bivectors must carry both representations $\bl{15}$ and $\widetilde{\bl{15}}$.

A spin off of the above reasoning is that the representations $\bl{15}$ and $\widetilde{\bl{15}}$ must be equivalent to each other. In fact, this is an immediate consequence of the fact that these two representations are related to each other by means of the hermitian conjugation:
$$ B_\alpha^{\;\;\beta} \overset{ \dagger }{\longrightarrow} (B^\dagger)^{\dot{\beta}}_{\;\;\dot{\alpha}} \,,$$
where Eq. (\ref{BarDagger}) has been used. However, what is more interesting about this equivalence is that it can be seen as a consequence of the map $\sim$. Let us prove this.
If $V^\mu$ and $W^\mu$ are vectors, their index structures in the spinorial formalism  are $V_{\alpha\dot{\beta}}$ and $W_{\alpha\dot{\beta}}$ or, equivalently, using the representation $\widetilde{\bl{6}}$ instead of $\bl{6}$, $\widetilde{V}^{\dot{\alpha}\beta}$ and $\widetilde{W}^{\dot{\alpha}\beta}$. The only natural way to generate the spinorial representation of bivector $(V\wedge W)^{\mu\nu}$, i.e. an object with the index structure $(V\wedge W)_\alpha^{\;\;\beta}$, that is bilinear and skew symmetric in $V$ and $W$ is the following:
\begin{equation}\label{B1}
  (V\wedge W)_\alpha^{\;\;\beta} = c\,\left( V_{\alpha \dot{\gamma}} \, \widetilde{W}^{\dot{\gamma} \beta} - W_{\alpha \dot{\gamma}} \, \widetilde{V}^{\dot{\gamma} \beta}\right) \,,
\end{equation}
with $c$ being some real constant. This constant is somehow arbitrary, so that henceforth we will set $c=1$. However, instead of using the representation $\bl{15}$ to write the bivector we could, equivalently, use the representation $\widetilde{\bl{15}}$, in which case we would have
\begin{equation}\label{B2}
  (\widetilde{V\wedge W})^{\dot{\alpha}}_{\;\;\dot{\beta}} = \left( \widetilde{V}^{\dot{\alpha} \gamma}  W_{\gamma \dot{\beta}} \,  -
\widetilde{W}^{\dot{\alpha} \gamma}  V_{\gamma \dot{\beta}}  \right) \,.
\end{equation}
Assuming that $V = \sigma_\mu$ and $W=\sigma_\nu$, for all possible values of $\mu,\nu$, we can check that these expressions for the spinorial representation of $(V\wedge W)^{\mu\nu}$ both have traces whose real parts vanish, as it should happen. Now, taking the hermitian conjugate of Eq. (\ref{B1}) we are led to
\begin{equation*}\label{B3}
  \left[(V\wedge W)^{\dagger}\right]^{\dot{\beta}}_{\;\;\dot{\alpha}}  =  \left( (\widetilde{W}^\dagger)^{ \dot{\beta} \gamma }  (V^\dagger)_{\gamma \dot{\alpha}} \, - (\widetilde{V}^\dagger)^{ \dot{\beta} \gamma }  (W^\dagger)_{\gamma \dot{\alpha}} \right) =
\left( \widetilde{W}^{ \dot{\beta} \gamma }  V_{\gamma \dot{\alpha}} \, - V^{ \dot{\beta} \gamma }  W_{\gamma \dot{\alpha}} \right) \,,
\end{equation*}
where in the last step it has been used that vectors are represented in the spinorial formalism by hermitian matrices. Now, comparing the latter equation with (\ref{B2}), we conclude that if the matrix $B$ transforms according to $\bl{15}$ and is the spinorial representation of a bivector $B^{\mu\nu}$, then the matrix $\widetilde{B}$ which carries the representation $\widetilde{\bl{15}}$ and represents the same bivector is related to $B^\dagger$ by the following relation:
\begin{equation}\label{B4}
  \widetilde{B} \,=\, -\, B^\dagger \,.
\end{equation}
This proves that the equivalence of the representations $\bl{15}$ and $\widetilde{\bl{15}}$, which comes from the hermitian conjugation, is associated to the map $\sim$ that connects the vector representations $\bl{6}$ and $\widetilde{\bl{6}}$. Note that the index contractions on the right hand side of Eqs. (\ref{B1}) and (\ref{B2}) are just the matrix multiplications, so that these equations can be elegantly written omitting the indices as follows:
\begin{equation}\label{B5}
  V\wedge W = V\widetilde{W} - W\widetilde{V} \quad \text{and} \quad \widetilde{V\wedge W} =  \widetilde{V}W - \widetilde{W}V \,.
\end{equation}

From the latter expression for the spinorial representation of bivectors, it is natural to guess that 3-vectors should be somehow related to the expressions
\begin{equation}\label{T1}
  V\wedge W \wedge Z = V\widetilde{W} Z - Z\widetilde{W} V + W\widetilde{Z}V -  V\widetilde{Z}W +  Z\widetilde{V} W - W\widetilde{V}Z \,.
\end{equation}
The index structure of the above object is $(V\wedge W \wedge Z)_{\alpha\dot{\beta}}$. Taking the hermitian conjugate of the above expression and using the fact that vectors are represented by hermitian matrices, we easily find that
\begin{equation}\label{T2}
  (V\wedge W \wedge Z)^\dagger = -(V\wedge W \wedge Z) \,.
\end{equation}
so that such object carries the representation $\bl{10}$, see Eqs. (\ref{AntiHermit}) and (\ref{TenRepIrred}). However, at this point a red light of caution might turn on, since in six dimensions a general 3-vector has 20 degrees of freedom while the representation $\bl{10}$ can only accommodate 10 degrees of freedom. The solution to this puzzle is that in six dimensions the Hodge dual operation maps 3-vectors into 3-vectors and canonically split the 20-dimensional space of 3-vectors into the direct sum of two 10-dimensional spaces, namely the space of self-dual 3-vectors and the space of anti-self-dual 3-vectors. Each of these subspaces carry an irreducible representation of the Lorentz group. Explicitly, we have that if $\bl{T}$ is some 3-vector and $\star \bl{T}$ is its Hodge dual then
$$  \underbrace{T^{\mu\nu\lambda}}_{20} = \underbrace{(T^+)^{\mu\nu\lambda}}_{10}  \,+\, \underbrace{(T^-)^{\mu\nu\lambda}}_{10} \;,\quad \text{ where } \bl{T}^{\pm} \equiv \frac{1}{2}  \left( \bl{T} \pm \star \bl{T} \right)  \,. $$
Using the fact that $\star(\star \bl{T}) = \bl{T}$, one can check that  $\bl{T}^+$ and $\bl{T}^-$ are self-dual  and anti-self-dual respectively, i.e. $\star \bl{T}^{\pm}=\pm \bl{T}^{\pm}$. Thus the spinorial representation given in Eq. (\ref{T1}) account just for half of the degrees of freedom of the 3-vector, let us say the self-dual part. The other half should be on the representation $\bl{10}'$, which is defined as the anti-hermitian part of the representation $\bar{\bl{2}}^{-1}\otimes \bl{2}^{-1}$ and is not equivalent to the representation $\bl{10}$. The representation  $\bl{10}'$ should account to the spinorial representation of the anti-self-dual part of 3-vector and the element of $M(2;\mathbb{H})$ that represents this part is given by
\begin{equation}\label{T3}
  V\wedge W \wedge Z = \widetilde{V}W\widetilde{Z}  - \widetilde{Z}W\widetilde{V}  + \widetilde{W}Z\widetilde{V} - \widetilde{V}Z\widetilde{W} +  \widetilde{Z}V\widetilde{W}  - \widetilde{W}V\widetilde{Z}   \,.
\end{equation}
Note, in particular, that the expression on the right hand side of the above equation has the index structure $(V\wedge W \wedge Z)^{\dot{\alpha}\beta}$ and can be checked to be an anti-hermitian matrix, in accordance with the assertion that it belongs to the representation $\bl{10}'$.

In Eqs. (\ref{T1}) and (\ref{T3}) we have used the same symbol, namely $V\wedge W \wedge Z$, to refer to two different and independent parts of the 3-vector $(V\wedge W \wedge Z)^{\mu\nu\lambda}$, which is certainly not appropriate. In order to correct this unsuitable notation, let us define the spinorial representation of the 3-vector $T^{\mu\nu\lambda}=(V\wedge W \wedge Z)^{\mu\nu\lambda}$ by the pair of matrices $(T^+,T^-)$ that carries the representation $\bl{10}\oplus\bl{10}'$ and is defined by:
\begin{equation}\label{T+-}
  \left\{
     \begin{array}{ll}
       (V\wedge W \wedge Z)^+ &=   V\widetilde{W} Z - Z\widetilde{W} V + W\widetilde{Z}V -  V\widetilde{Z}W +  Z\widetilde{V} W - W\widetilde{V}Z \,,\\
       (V\wedge W \wedge Z)^- &= \widetilde{V}W\widetilde{Z}  - \widetilde{Z}W\widetilde{V}  + \widetilde{W}Z\widetilde{V} - \widetilde{V}Z\widetilde{W} +  \widetilde{Z}V\widetilde{W}  - \widetilde{W}V\widetilde{Z} \,.
     \end{array}
   \right.
\end{equation}
A question that can be raised at this point is how can we prove that the representation $\bl{10}$ is connected to the self-dual part of the 3-vectors instead of the anti-self-dual part. The answer is that the association of $T^+$ to $\bl{10}$ and $T^-$ to $\bl{10}'$ is arbitrary, we could do the other way around and say that $T^+$ is in the representation $\bl{10}'$. But once we make a choice we must stick to it. The reason for the arbitrariness is that the volume-form, which is used to define the Hodge dual, is well defined only up to a sign, and a change in this sign swap the notions of self-dual and anti-self-dual 3-forms. In this article we will make the choice of associating the self-dual part to $\bl{10}$, as done in Eq. (\ref{T+-}).

Aiming to give a concrete realization of the abstract concepts introduced in the present section, we end up with a few explicit matrix representations of the bivectors and 3-vectors. Denoting by $\{E_\mu\}$ the Lorentz frame adopted in the expansion of our vectors throughout the text, i.e. assuming that the spinorial representation of $E_\mu$ is $\sigma_\mu$ (using the representation $\bl{6}$), it follows that the spinorial representation of the bivector $(E_1\wedge E_2)^{\mu\nu}$ is given by
$$ (E_1\wedge E_2)_{\alpha}^{\;\;\beta} = \sigma_1 \widetilde{\sigma}_2 - \sigma_2 \widetilde{\sigma}_1
 = \sigma_1  \sigma_2 - \sigma_2 \sigma_1 =
\left[
  \begin{array}{cc}
    -2\bl{K} & 0 \\
    0 & -2\bl{K} \\
  \end{array}
\right]\,.
  $$
Note that the trace of the latter matrix has vanishing real part, as it should. Analogously, after some straightforward algebra, one can check that the spinorial representation of the 3-vectors $(E_1\wedge E_2 \wedge E_3)^{\mu\nu\lambda}$ and $(E_0\wedge E_1 \wedge E_2)^{\mu\nu\lambda}$ are respectively given by:
$$ (E_1\wedge E_2 \wedge E_3)^+_{\alpha\dot{\beta}} =  \left[
  \begin{array}{cc}
    0 & 6 \\
    -6 & 0 \\
  \end{array}
\right] \;,\;\;
(E_1\wedge E_2 \wedge E_3)^{-\;\dot{\alpha}\beta} =  \left[
  \begin{array}{cc}
    0 & 6 \\
    -6 & 0 \\
  \end{array}
\right]\;, \;\text{and} $$
$$ (E_0\wedge E_1 \wedge E_2)^+_{\alpha\dot{\beta}} =  \left[
  \begin{array}{cc}
    -6\bl{K} & 0 \\
    0 & -6\bl{K} \\
  \end{array}
\right] \;,\;\;
(E_0\wedge E_1 \wedge E_2)^{-\;\dot{\alpha}\beta} =  \left[
  \begin{array}{cc}
    6\bl{K} & 0 \\
    0 & 6\bl{K} \\
  \end{array}
\right]\;. $$
Note that each 3-vector is represented by a pair of anti-hermitian matrices.

\section{A brief review of the $SU(4)$ spinor formalism}\label{Sec.ReviewSU4}

As mentioned in the introduction section, the chiral spinors of six-dimensional Minkowski space can also be written in terms of four-component objects over the complex field, instead of using two-component quaternionic objects. Such approach comes from the fact that the Spin group of $\mathbb{R}^6$ is $SU(4)$ and by means of complexifying the space $\mathbb{R}^6$ we can handle the case of Lorentz signature and, therefore, represent chiral spinors as objects with four complex components, $\psi^A$ (positive chirality) and $\phi_A$ (negative chirality) with $A,B,C\in{1,2,3,4}$. If $U^A_{\;\;\,B}$ is an element of $SU(4)$, which can be associated to some Lorentz transformation, then these spinors transform as follows:
$$ \psi^A \overset{U}{\longrightarrow}  U^A_{\;\;\,B}\psi^B \quad \text{ and }\quad  \phi_A \overset{U}{\longrightarrow} [(U^{-1})^t]_A^{\;\;\,B}\phi_B \,,  $$
In particular, one can check that $\psi^A \phi_A$ is invariant by the action of the spin group. A Dirac spinor is then given by a pair of chiral spinors of different chiralities, namely $\Psi=(\psi^A,\phi_A)$.

In this $SU(4)$ formalism, vectors are represented by objects with the index structure $V^{AB}=V^{[AB]}$ or, equivalently, by $\tilde{V}_{AB}=\tilde{V}_{[AB]}$. Indeed, as a check, note that these skew symmetric objects have 6 degrees of freedom, since the indices $A$ and $B$ can assume four possible values. The connection between these two forms of representing a vector is given by the Levi-Civita symbol $\varepsilon_{ABCD}$. More precisely, we have
$$ \tilde{V}_{AB} = \frac{1}{2} \varepsilon_{ABCD} V^{CD} \,.$$
The inner product of two vectors $V^\mu$ and $W^\mu$ is given in this spinorial formalism by $V^{AB} \tilde{W}_{AB}$. Since the matrices of $SU(4)$ have unit determinant, it follows that the symbol $\varepsilon_{ABCD}$ is invariant by the action of the spin group, just as the Minkowski metric is invariant by the action of the Lorentz group.

Bivectors are represented by objects with the index structure $B^A_{\;\;\,B}$ with vanishing trace, i.e. $B^A_{\;\;\,A}=0$. Such objects have 15 independent components and carry an irreducible representation of $SU(4)$, in accordance with the fact that in six dimensions the space of bivectors is 15-dimensional and carries an irreducible representation of the Lorentz group. In turn, 3-vectors are represented by a pair of symmetric objects, $(T^{AB},T_{AB})$, where $T^{AB}=T^{(AB)}$ and $T_{AB}= T_{(AB)}$. A 3-vector such that $T_{AB}$ vanishes is said to be self-dual, whereas a 3-vector of the form $(0,T_{AB})$ is anti-self-dual.

Since a general chiral spinor has four components, a frame of spinors of positive chirality must be comprised by four elements, let us denote these spinors by $\{\chi_1^A, \chi_2^A,\chi_3^A,\chi_4^A \} = \{\chi_{\mf{a}}^A\}$, where $\mf{a},\mf{b} \in \{1,2,3,4\}$ label the spinors of the frame. Thus, a general spinor of positive chirality can be spanned as
$$\psi^A = c^{\mf{a}}\,\chi_{\mf{a}}^A \,,$$
where $c^{\mf{a}}$ are complex constants. In the same fashion, we can introduce a frame of spinors of negative chirality $\{\zeta^{\mf{a}}_A\}$, so that a general spinor of negative chirality can be written as
$$\phi_A = c_{\mf{a}}\,\zeta^{\mf{a}}_A \,. $$
We shall choose the latter frame as being dual to the former, i.e. we shall assume that $\chi_{\mf{a}}^A \zeta^{\mf{b}}_A = \delta_{\mf{a}}^{\;\;\mf{b}}$.
By forming skew symmetric products of the frame elements we can then generate a frame of vectors:
$$ e_1^{\,AB} \eq \chi_{[1}^{\,A}\chi_{2]}^{\,B} \quad,\quad e_2^{\,AB} \eq \chi_{[1}^{\,A}\chi_{3]}^{\,B}  \quad,\quad
 e_3^{\,AB} \eq \chi_{[1}^{\,A}\chi_{4]}^{\,B} \,, $$
\begin{equation}\label{NullFrame1}
  \theta^{1\,AB} \eq \chi_{[3}^{\,A}\chi_{4]}^{\,B}  \quad,\quad \theta^{2\,AB}  \eq \chi_{[4}^{\,A}\chi_{2]}^{\,B} \quad,\quad
\theta^{3\,AB} \eq \chi_{[2}^{\,A}\chi_{3]}^{\,B}  \,.
\end{equation}
This is a, so called, null frame, since all vectors are null. In fact, it is simple matter to prove that any vector that is the skew symmetric product of two spinors, $V^{AB}= \chi^{[A} \xi^{B]}$ is light-like. More precisely, the only nonvanishing inner products of the latter vector frame are
\begin{equation}\label{InnerProdNullFrame1}
  e_i^{\;\mu} \,\theta^j_{\;\mu} = e_i^{\;AB} \,\tilde{\theta}^j_{\;AB} = \frac{1}{2}\delta_i^{\;j}\,.
\end{equation}
We can also lower the pair of spinorial indices of the vector frame and then write the result in terms of the frame of spinors of negative chirality, with the final result being given by
$$ \tilde{e}_{1\,AB} \eq \zeta^{[3}_{\,A}\zeta^{4]}_{\,B} \quad,\quad \tilde{e}_{2\,AB} \eq \zeta^{[4}_{\,A}\zeta^{2]}_{\,B} \quad,\quad
 \tilde{e}_{3\,AB} \eq \zeta^{[2}_{\,A}\zeta^{3]}_{\,B} $$
\begin{equation}\label{NullFrame2}
  \tilde{\theta}^{1}_{\,AB} \eq \zeta^{[1}_{\,A}\zeta^{2]}_{\,B} \quad,\quad \tilde{\theta}^{2}_{\,AB}\eq \zeta^{[1}_{\,A}\zeta^{3]}_{\,B} \quad,\quad
\tilde{\theta}^{3}_{\,AB} \eq \zeta^{[1}_{\,A}\zeta^{4]}_{\,B} \,.
\end{equation}
At this point it is worth stressing that although we can lower or raise a pair of skew symmetric spinorial indices by means of the Levi Civita symbol, there is no canonical way of converting a lower spinorial index into an upper spinorial index, i.e. the objects $\psi^A$ and $\phi_A$ carry independent representations of $SU(4)$, just as the representations $\bl{2}$ and $\bar{\bl{2}}^{-1}$ are inequivalent representations of $SL(2;\mathbb{H})$.

Bivectors and 3-vectors can also be written using these frames of spinors. For instance, the bivectors $(e_1\wedge e_2)^{\mu\nu}$ and $(e_1\wedge \theta^1)^{\mu\nu}$ have the following spinorial representations in the $SU(4)$ spinorial formalism \cite{CaBru,Bat-Book}\footnote{Here we have multiplied the bivector representations $B^A_{\;\;B}$ adopted in Refs. \cite{CaBru,Bat-Book} by a global factor of $-4$. This global factor was arbitrarily chosen in these references and here it is conveniently redefined. }:
\begin{equation}\label{BivectorsSU4}
  \left.
  \begin{array}{ll}
    (e_1\wedge e_2)^A_{\;\;\,B} &=  \chi_1^{A}\,\zeta^4_{B} \,, \\
\quad \\
    (e_1\wedge \theta^1)^A_{\;\;\,B} &= \frac{1}{2}\left(  \chi_1^{A}\,\zeta^1_{B} + \chi_2^{A}\,\zeta^2_{B}
- \chi_3^{A}\,\zeta^3_{B} -\chi_4^{A}\,\zeta^4_{B}\right)\,.
  \end{array}
\right.
\end{equation}
In the same vein, the 3-vectors $(e_1\wedge e_2 \wedge e_3)^{\mu\nu\lambda}$ and $(e_1\wedge e_2 \wedge \theta^2)^{\mu\nu\lambda}$ are given in the spinorial formalism by\footnote{Again, the definitions of $T^{+\,AB}$ and $T^-_{AB}$ differ from the ones adopted in Ref. \cite{Bat-Book} by a global multiplicative factor of $-3$, which is of no fundamental relevance.}:
\begin{equation}\label{3-vectorsSU4}
  \left.
  \begin{array}{ll}
    (e_1\wedge e_2 \wedge e_3)^{+\,AB} &=  -3\chi_1^{A}\,\chi_1^{B} \;,\;\;  (e_1\wedge e_2 \wedge e_3)^{-}_{\,AB} = 0 \,, \\
\quad \\
    (e_1\wedge e_2 \wedge \theta^2)^{+\,AB} &=  3\chi_{(1}^{A}\,\chi_{2)}^{B} \;,\;\;
 (e_1\wedge e_2 \wedge \theta^2)^{-}_{\,AB} = -3 \zeta^{(1}_{A} \zeta^{4)}_{B} \,.
  \end{array}
\right.
\end{equation}

These basic tools of the $SU(4)$ spinorial formalism are enough in order to make the bridge between this approach and the $SL(2;\mathbb{H})$ spinorial formalism, a task that will be tackled in the sequel. For more details on the $SU(4)$ spinorial approach, the reader is referred to \cite{CaBru,Carlos,Weinberg-Conf6D,Mason6D,Batista_Conf6D,Kerr6D, Koller83}.

\section{Constructing vectors from a pair of spinors}\label{Sec.VectorFromSpinors}

Given a pair of spinors of the same chirality it is simple matter to generate a vector out of them. For instance, if $\psi_\alpha$ and $\phi_\beta$ are spinors of positive chirality then it follows that 
$$ V_{\alpha \dot{\beta}} = \psi_\alpha\,\bar{\phi}_{\dot{\beta}} + \phi_{\alpha}\, \bar{\psi}_{\dot{\beta}}  $$
is a vector, since it is a $2\times 2$ hermitian matrix carrying the representation $\bl{2}\otimes \bar{\bl{2}}$, i.e. it is on the representation $\bl{6}$. Note how the index notation makes easy to analyse which combinations of spinors can yield an object carrying the correct group representation. Omitting indices, the latter vector is written as
\begin{equation}\label{VBaez}
  V = \Psi\,\Phi^\dagger + \Phi\,\Psi^\dagger\,.
\end{equation}
Likewise, given two spinors of negative chirality we can generate an object carrying the representation $\widetilde{\bl{6}}$. The vector (\ref{VBaez}) is the vector defined in Ref. \cite{Baez:2010bbj} using a pair of spinors, when discussing about two component spinors in dimensions 4, 6 and 10. In the same fashion, using the $SU(4)$ spinorial approach, the spinors of positive chirality have index structures given by $\psi^A$ and $\phi^A$, so that we can easily form the following vector out of them: $\psi^{[A} \phi^{B]}$. However, as a matter of fact, it turns out that the latter vector constructed in the $SU(4)$ formalism is not the one built in Eq. (\ref{VBaez}). Indeed, while $\psi^{[A} \phi^{B]}$ is always a null vector (since in the $SU(4)$ formalism a vector is light-like if, and only if, it can be written as $\psi^{[A} \phi^{B]}$ for some pair of spinors $\psi^A$ and $\phi^A$), the same is not true for the one given in Eq. (\ref{VBaez}). The aim of the present section is then to show how one can build the null vector  $\psi^{[A} \phi^{B]}$ in the quaternionic formalism.

Now let us return to the quaternionic spinorial formalism. Consider a pair of chiral spinors $\bl{\Psi}$ and $\bl{\Phi}$, let us say carrying the representation $\bl{2}$. Then, just as happens in the $SU(4)$ spinorial formalism, we must be able to construct a vector $V_{[\Psi,\Phi]}$ with the following properties:
$$ \left\{
     \begin{array}{ll}
       (i)&\text{$V_{[\Psi,\Phi]}$ is bilinear in $\bl{\Psi}$ and $\bl{\Phi}$}\,, \\
       (ii)&  \text{$V_{[\Psi,\Phi]} = - V_{[\Psi,\Phi]}$ }\,, \\
       (iii)& \text{$V_{[\Psi,\Phi]}$ is an hermitian element of $M(2;\mathbb{H})$} \,,\\
       (iv)&\text{det$(V_{[\Psi,\Phi]})$ must vanish}\,. \\
          \end{array}
   \right.
$$
The constraints $(i)$ and $(ii)$ comme from the fact that they hold in the $SU(4)$ formalism, since $V_{[\psi,\phi]}^{AB} = \psi^{[A} \phi^{B]}$, while properties $(iii)$ and $(iv)$ follow from the fact that in the quaternionic formalism vectors are represented by hermitian matrices and that light-like vectors (which is the case of $V_{[\psi,\phi]}^{AB} = \psi^{[A} \phi^{B]}$) are represented by matrices with vanishing determinant.

The most general way of building an hermitian matrix that is bilinear in $\bl{\Psi}$ and $\bl{\Phi}$ is given by
\begin{equation}\label{V0}
  V_{[\Psi,\Phi]} = \frac{1}{2}\, F \,\left( \bl{\Psi} \, G \,\bl{\Phi}^\dagger \right)\, H +
\frac{1}{2}\,H^\dagger\,\left(\bl{\Phi}\, G^\dagger\,\bl{\Psi}^\dagger\right) \,F^\dagger \,
\end{equation}
with $F$, $G$, and $H$ being fixed $2\times 2$ quaternionic matrices. Then, properties $(i)$ and $(iii)$ are automatically satisfied once the expression (\ref{V1}) is assumed for $V_{[\Psi,\Phi]}$. The global factor of $1/2$ in the above expression could obviously have been absorbed in $G$, but we made it explicit for future convenience. It is simple to check that any element of $M(2;\mathbb{H})$ can be written as a real number times an $SL(2;\mathbb{H})$ matrix. Doing this for the matrices $F$ and $H$ and absorbing the real multiplicative factors into $G$ it follows that we can, without loss of generality, assume that $F$ and $H$ are both elements of $SL(2;\mathbb{H})$. Moreover, choosing $H$ and $G$ to be such that
$$  H = F^\dagger \quad \text{ and } \quad   G^\dagger = -G $$
it follows that
\begin{equation}\label{V01}
  V_{[\Psi,\Phi]} = \frac{1}{2}\, F \,\left( \bl{\Psi} \, G \,\bl{\Phi}^\dagger  - \bl{\Phi} \, G \,\bl{\Psi}^\dagger \right)\, F^\dagger  \,,
\end{equation}
which automatically obeys the desired property $(ii)$. In addition, defining $ \bl{\Psi}'  = F  \bl{\Psi}$ and  $\bl{\Phi}'  = F  \bl{\Phi}$, which amounts to a Lorentz transformation on the spinors, it follows that the latter expression can be written as
\begin{equation}\label{V02}
  V_{[\Psi,\Phi]} =    \frac{1}{2}\,  \left(\bl{\Psi}' \, G \,\bl{\Phi}^{'\,\dagger}  - \bl{\Phi}' \, G \,\bl{\Psi}^{'\,\dagger}\right) \,.
\end{equation}
This proves that by means of choosing a convenient Lorentz frame we can always get rid of  the matrix $F$ in Eq. (\ref{V01}). Hence, for convenience, we shall assume $F = \mathbb{I}_2$. It then remains to impose the condition $(iv)$, namely that the matrix $V_{[\Psi,\Phi]}$ have vanishing determinant, which will constrain the matrix $G$. As we shall see in the sequel, it turns out that a solution is still possible in the particular case
$G= \bl{g} \mathbb{I}_2$, where $\bl{g}$ is some quaternionic number. Therefore, henceforth we shall adopt the following expression for $V_{[\Psi,\Phi]}$:
\begin{equation}\label{V1}
  V_{[\Psi,\Phi]} =  \frac{1}{2}\,  \left( \bl{\Psi} \, \bl{g} \,\bl{\Phi}^\dagger  -  \bl{\Phi}\, \bl{g}\,\bl{\Psi}^\dagger \right) \,.
\end{equation}
The property $G^\dagger = -G$ now translates to the condition that $\bl{g}$ is purely quaternionic, namely $\bar{\bl{g}} = - \bl{g}$.

Now, defining the components of the spinors $\bl{\Psi}$ and $\bl{\Phi}$ to be
$$ \bl{\Psi}= \left[
           \begin{array}{c}
             \bl{a} \\
             \bl{b} \\
           \end{array}
         \right]\quad
\text{ and } \quad
\bl{\Phi} = \left[
           \begin{array}{c}
             \bl{c} \\
             \bl{d} \\
           \end{array}
         \right]\,,
 $$
where $\bl{a},\bl{b},\bl{c}$ and $\bl{d}$ are quaternionic numbers, it follows that the determinant of $V_{[\Psi,\Phi]}$ is given by
\begin{align}
  \text{det}(V_{[\Psi,\Phi]}) = \frac{1}{4}( \bl{a} \bl{g} \bar{\bl{c}} - \bl{c} \bl{g} \bar{\bl{a}}) ( \bl{b} \bl{g} \bar{\bl{d}}  - \bl{d} \bl{g} \bar{\bl{b}}) - \frac{1}{4}( \bl{b} \bl{g} \bar{\bl{c}} - \bl{d} \bl{g} \bar{\bl{a}} ) ( \bl{a} \bl{g} \bar{\bl{d}} -  \bl{c} \bl{g} \bar{\bl{b}})   \,.
\end{align}
Distributing the products in the latter equation, we end up with eight terms and just one of them is quadratic in $\bl{a}$, which means that this term should vanish independently from the others, since such determinant must vanish for arbitrary values of $\bl{a}$, $\bl{b}$, $\bl{c}$ and $\bl{d}$. The mentioned term is given by $\frac{1}{4}\bl{g}^2 |\bl{a}|^2 |\bl{d}|^2 $, so that we conclude that $\bl{g}^2$ should vanish. Interestingly, assuming $\bl{g}^2=0$, along with the previous constraint $\bar{\bl{g}}=-\bl{g}$, it can be proved that the identity
\begin{equation}\label{Idabc}
  \bl{g}\,\bl{a}\,\bar{\bl{b}}\,\bl{g}\,\bl{c} + \bl{g}\,\bl{b}\,\bar{\bl{c}}\,\bl{g}\,\bl{a} -\bl{g}\,\bl{a}\,\bar{\bl{c}}\,\bl{g}\,\bl{b} =0 \,,
\end{equation}
holds for any $\bl{a},\bl{b},\bl{c}\in \mathbb{H}$. This identity, in turn, implies that det$(V_{[\Psi,\Phi]})$ vanishes. Therefore, $V_{[\Psi,\Phi]}$ given in Eq. (\ref{V1}) obeys all the desired properties $(i)-(iv)$ as long as $\bl{g}$ obeys to the following constraints:
\begin{equation}\label{gcontraints}
   \bar{\bl{g}} = - \bl{g} \quad \text{and} \quad  \bl{g}^2 = 0\,.
\end{equation}
In particular, this implies that $\bar{\bl{g}} \bl{g}=0$, whose unique solution on the quaternions is $\bl{g}=0$. However, working in the complexified quaternions there exist nontrivial solutions for this constraint, as exemplified by $\bl{g}= 1 + \bl{iJ}$.  This is just a particular solution for the constraints (\ref{gcontraints}), with the general solution being
\begin{equation}\label{g}
  \bl{g} = g_1\bl{I} + g_2\,\bl{J} + \bl{i} \sqrt{g_1^2 + g_2^2}\,\,\bl{K} \,,
\end{equation}
where $g_1$ and $g_2$ are arbitrary complex numbers with respect to the imaginary unit $\bl{i}$. Thus, our final expression for $V_{[\Psi,\Phi]}$ is given by Eq. (\ref{V1}) with $\bl{g}$ being given by (\ref{g}).

Note that here we have supposed that the spinors used to build the vector have positive chirality, i.e. carry the representation $\bl{2}$. Nevertheless, in the $SU(4)$ formalism we can also use a pair of negative chirality spinors to build a null vector. Likewise, if $\bl{\Psi}$ and $\bl{\Phi}$ are now are assumed to be spinors of negative chirality in the $SL(2;\mathbb{H})$ formalism, i.e. spinors that carry the representation $\bar{\bl{2}}^{-1}$, then the same expression (\ref{V1}) can also be used to build a vector. In this case, the object $V_{[\Psi,\Phi]}$ will be on the representation $\widetilde{\bl{6}}$.




\section{The bridge between the $SL(2;\mathbb{H})$ and the $SU(4)$ spinors}\label{Sec.Bridge}

It turns out that a quaternionic degree of freedom can be consistently encoded in a pair of complex numbers. More explicitly, given a quaternion
$$ \bl{a} = a^{\mf{m}}\bl{q}_{\mf{m}} = a^1 \bl{I} + a^2 \bl{J} + a^3 \bl{K}  + a^4 \,,$$
it can be conveniently written as
\begin{equation}\label{HCC}
   \bl{a} = (a^4 + a^1 \bl{I} )+ (a^2 + a^3 \bl{I} )\,\bl{J} = \bl{z}^1 + \bl{z}^2 \,\bl{J}\,,
\end{equation}
where $\bl{z}^1$ and $\bl{z}^2$ are the ``complex'' numbers given by $\bl{z}^1= a^4 + a^1 \bl{I}$ and  $\bl{z}^2= a^2 + a^3 \bl{I}$. Thus, knowing the pair of  ``complex'' numbers $(\bl{z}^1,\bl{z}^2)$ is equivalent to knowing the quaternion $\bl{a}$.  This simple observation is the key to connect the two-component quaternionic spinorial formalism to the four-component $SU(4)$ spinorial formalism. Before proceeding, let us mention that here by a complex number (without quotes) it always meant one with respect to the imaginary unit $\bl{i}$, like $z= r_1 + \bl{i} r_2$, where $r_1$ and $r_2$ are real numbers. On the other hand, the denomination ``complex'' (under quotes) refer to the case in which the imaginary unit is assumed to be $\bl{I}$, so $\bl{z} = r_1 + \bl{I} r_2$ is a ``complex'' number.

Using the decomposition (\ref{HCC}) in the $SL(2;\mathbb{H})$ formalism, it follows that a general spinor on the representation $\bl{2}$ can be written as
$$ \Psi = \left[
            \begin{array}{c}
              \bl{z}^1 + \bl{z}^2 \bl{J} \\
              \bl{z}^3 + \bl{z}^4 \bl{J}  \\
            \end{array}
          \right] = \bl{z}^{\mf{a}} \Psi_{\mf{a}}\,,
 $$
where $\bl{z}^{\mf{a}}$ are ``complex'' numbers with respect to the imaginary unit $\bl{I}$ and $\Psi_{\mf{a}}$ is the following frame:
$$ \Psi_{1} = \left[
            \begin{array}{c}
              \bl{1} \\
              0  \\
            \end{array}
          \right]\;,\quad
\Psi_{2} = \left[
            \begin{array}{c}
              \bl{J} \\
              0  \\
            \end{array}
          \right]\;,\quad
\Psi_{3} = \left[
            \begin{array}{c}
              0 \\
              \bl{1}  \\
            \end{array}
          \right]\;,\quad
\Psi_{4} = \left[
            \begin{array}{c}
              0 \\
              \bl{J}  \\
            \end{array}
          \right]\,.$$
This shows how two quaternionic components can be associated to four complex numbers. However, instead of using $\{\Psi_{\mf{a}}\}$ as a frame for the for the $SL(2;\mathbb{H})$ spinors, it is more convenient for our purpose to adopt the frame $\{\chi_\mf{a}\}$ defined by:
$$ \chi_{1} = \frac{\sqrt{2}}{\sqrt{g_0 - g_1}}\left[
            \begin{array}{c}
              \bl{i\,s} \\
              0  \\
            \end{array}
          \right]\;,\quad
\chi_{2} = \frac{\sqrt{2}}{\sqrt{g_0 + g_1}} \left[
            \begin{array}{c}
              \bar{\bl{s}} \\
              0  \\
            \end{array}
          \right]\;,\quad
$$
$$
\chi_{3} = \frac{\sqrt{2}}{\sqrt{g_0 + g_1}} \left[
            \begin{array}{c}
              0 \\
              \bl{i}\,\bar{{\bl{s}}}  \\
            \end{array}
          \right]\;,\quad
\chi_{4} = \frac{\sqrt{2}}{\sqrt{g_0 - g_1}} \left[
            \begin{array}{c}
              0 \\
              \bl{s}  \\
            \end{array}
          \right]\,,$$
where
$$ \bl{s}\equiv\frac{1}{2}(1+\bl{i}\bl{J}) \quad \text{and} \quad g_0 \equiv \sqrt{g_1^2 + g_2^2} \,. $$
In this frame, a general spinor on the representation $\bl{2}$ is written as $\Psi = \bl{w}^{\mf{a}} \,\chi_{\mf{a}}$. The reason for using the frame $\{\chi_{\mf{a}}\}$, which uses the imaginary unit $\bl{i}$ in its definition, instead of using the simpler frame $\{\Psi_{\mf{a}}\}$ is that, as previously explained, the $SL(2;\mathbb{H})$ approach is naturally designed for the Lorentzian signature, whereas the $SU(4)$ spinorial formalism is related to the Euclidean signature, with the Lorentzian case obtained only after a complexification.

The latter frame is the bridge that links the representation $\bl{2}$ of $SL(2;\mathbb{H})$ to the representation carried by the spinors of positive chirality of $SU(4)$. Indeed, the spinors $\chi_{\mf{a}}$ of $SL(2;\mathbb{H})$ should be identified with the spinors $\chi_{\mf{a}}^{\,A}$ of $SU(4)$, that have been introduced in Sec. \ref{Sec.ReviewSU4}. In particular, now we can construct the null frame of vectors, introduced in Eq. (\ref{NullFrame1}), in terms of the $SL(2;\mathbb{H})$ spinors by means of Eq. \eqref{V1}. In order to do so, it is handy to use to the identities
\begin{align*}
  \bl{s}\,\bl{g}\,\bl{s} = \frac{g_2}{2} (\bl{J} -\bl{i}) \;,&\;\;
\bar{\bl{s}}\,\bl{g}\,\bar{\bl{s}} = \frac{g_2}{2} (\bl{J} +\bl{i}) \;, \\
       \bar{\bl{s}}\,\bl{g}\,\bl{s} = \frac{g_0+g_1}{2}\,(\bl{I} + \bl{i\,K})
\;,&\;\;   \bl{s}\,\bl{g}\,\bar{\bl{s}} = \frac{g_1-g_0}{2}\,(\bl{I} - \bl{i\,K})\;,
\end{align*}
which can be easily checked. The quaternion $\bl{s}$ also obeys the interesting identities $\bl{s}\bar{\bl{s}}=0$ and $\bl{s}^n = \bl{s}$. Thus, for instance, the null vector $e_1^\mu$, whose spinorial representation in the $SU(4)$ formalism is given by $\chi_{[1}^{A}\chi_{2]}^B$, is built in the $SL(2;\mathbb{H})$ approach by means of Eq. \eqref{V1}, which yields
$$  e_1 = V_{[\chi_1,\chi_2]} = \frac{1}{2}\left( \chi_1 \,\bl{g}\,\chi_2^\dagger - \chi_2 \,\bl{g}\,\chi_1^\dagger \right) =
\chi_{[1}\, \bl{g}\, \chi_{2]}^\dagger = \frac{1}{2}(\sigma_5 + \sigma_0) \,.$$
Likewise, we can obtain the matrices that represent the other elements of the frame. The final result is given by
\begin{equation*}
  \left.
     \begin{array}{ll}
       e_1 = \chi_{[1}\bl{g}\chi^\dagger_{2]} = \frac{1}{2}(\sigma_5 + \sigma_0)\;,\;\;
e_2 = \chi_{[1}\bl{g}\chi^\dagger_{3]} = \frac{1}{2}(\sigma_2 + \bl{i}\sigma_4)\;,\;\;
e_3 = \chi_{[1}\bl{g}\chi^\dagger_{4]} = \frac{1}{2}(\sigma_3+ \bl{i}\sigma_1)\;, \\
\quad \\
       \theta^1 = \chi_{[3}\bl{g}\chi^\dagger_{4]} = \frac{1}{2}(\sigma_5 - \sigma_0)\;,\;\;
\theta^2 = \chi_{[4}\bl{g}\chi^\dagger_{2]} = \frac{1}{2}(\sigma_2 - \bl{i}\sigma_4)\;,\;\;
\theta^3 = \chi_{[2}\bl{g}\chi^\dagger_{3]} = \frac{1}{2}(\sigma_3 - \bl{i}\sigma_1)\;.
     \end{array}
   \right.
\end{equation*}
In particular, we can check that all vectors of the frame $\{e_i,\theta^i\}$ are null and that the only nonvanishing inner products are
$$ e_i^{\;\mu} \, \theta^j_{\;\mu} = \frac{1}{2} \,\delta_i^{\;j} \,,$$
in perfect accordance with Eq. (\ref{InnerProdNullFrame1}). Note also that the latter frame obeys the following reality conditions:
$$ e_1^\star = e_1\;,\;\; \theta^{1\,\star} = \theta^1  \;,\;\; e_2^\star = \theta^2  \;,\;\; e_3^\star = \theta^3 \,,   $$
which means that the metric has Lorentzian signature, as it should.

We have defined the frame for the positive chirality spinors, $\{\chi_{\mf{a}}\}$, which carry the representation $\bl{2}$ and are the analogous of the spinors $\chi_{\mf{a}}^{\;A}$ in the $SU(4)$ formalism. Now, it remains to introduce the frame of negative chirality spinors, which is the $SL(2;\mathbb{H})$ analogue of the spinors $\zeta^{\mf{a}}_{\;A}$ of $SU(4)$. Aiming that each vector $\widetilde{V}$, carrying the representation $\widetilde{\bl{6}}$ in the $SL(2;\mathbb{H})$ formalism, should be associated to the corresponding vector with down indices in the $SU(4)$ approach, namely $\tilde{V}_{AB}$, it follows that the suitable choice of basis is
$$ \zeta^{1} = \frac{\sqrt{2}}{\sqrt{g_0 + g_1}}\left[
            \begin{array}{c}
             \bar{\bl{s}}  \\
              0  \\
            \end{array}
          \right]\;,\quad
\zeta^{2} = \frac{\sqrt{2}}{\sqrt{g_0 - g_1}} \left[
            \begin{array}{c}
              -\bl{i}\bl{s}  \\
              0  \\
            \end{array}
          \right]\;,\quad
$$
$$
\zeta^{3} = \frac{\sqrt{2}}{\sqrt{g_0 - g_1}} \left[
            \begin{array}{c}
              0 \\
              -\bl{s} \\
            \end{array}
          \right]\;,\quad
\zeta^{4} = \frac{\sqrt{2}}{\sqrt{g_0 + g_1}} \left[
            \begin{array}{c}
              0 \\
               	\bl{i}\bar{\bl{s}}   \\
            \end{array}
          \right]\,.$$
Indeed, with this choice of frame we obtain the following vector frame that is the $SL(2;\mathbb{H})$ analogous of the frame (\ref{NullFrame2}) defined in the $SU(4)$ formalism:
\begin{equation*}
  \left.
     \begin{array}{ll}
       \widetilde{e}_1 = \zeta^{[3}\bl{g}\zeta^{4]\dagger} = \frac{1}{2}(\widetilde{\sigma}_5 + \widetilde{\sigma}_0)\;,\;\;
\widetilde{e}_2 = \zeta^{[4}\bl{g}\zeta^{2]\dagger} = \frac{1}{2}(\widetilde{\sigma}_2 + \bl{i}\widetilde{\sigma}_4)\;,\;\;
\widetilde{e}_3 = \zeta^{[2}\bl{g}\zeta^{3]\dagger} = \frac{1}{2}(\widetilde{\sigma}_3+ \bl{i}\widetilde{\sigma}_1)\;, \\
\quad \\
       \widetilde{\theta}^1 = \zeta^{[1}\bl{g}\zeta^{2]\dagger} = \frac{1}{2}(\widetilde{\sigma}_5 - \widetilde{\sigma}_0)\;,\;\;
\widetilde{\theta}^2 = \zeta^{[1}\bl{g}\zeta^{3]\dagger} = \frac{1}{2}(\widetilde{\sigma}_2 - \bl{i}\widetilde{\sigma}_4)\;,\;\;
\widetilde{\theta}^3 = \zeta^{[1}\bl{g}\zeta^{4]\dagger} = \frac{1}{2}(\widetilde{\sigma}_3 - \bl{i}\widetilde{\sigma}_1)\;.
     \end{array}
   \right.
\end{equation*}

Let us compare the expressions for the frame vectors in the $SL(2;\mathbb{H})$ and $SU(4)$ spinorial formalisms:
\begin{equation*}
  \left\{
     \begin{array}{ll}
       SU(4):\;\; V^{AB} = \chi_{[\mf{a}}^{\;A}\,\chi_{\mf{b}]}^{\;B} \quad \text{ and }  \quad
 \tilde{V}_{AB} = \zeta^{[\mf{a}}_{\;A}\,\zeta^{\mf{b}]}_{\;B} \,, \\
\quad \\
       SL(2;\mathbb{H}):\;\; V = \chi_{[\mf{a}} \,\bl{g}\,\chi_{\mf{b}]}^{\dagger} \quad \text{ and }  \quad
 \widetilde{V} = \zeta^{[\mf{a}}\,\bl{g}\,\zeta^{\mf{b}]\dagger}\,.
     \end{array}
   \right.
\end{equation*}
Note that the expressions in the two approaches are almost identical, with the only differences being that in the $SL(2;\mathbb{H})$ formalism we need to put the quaternion $\bl{g}$ between the spinors and we must take the hermitian conjugation of the spinor on the right hand side of the matrix multiplication. The latter fact is of great technical importance, since a vector is represented by a $2\times 2$ matrix in the $SL(2;\mathbb{H})$ approach and in order to generate such a matrix we need to multiply a column matrix on the left to a row matrix on the right, reason why the hermitian conjugation is essential. It turns out that the resemblance between these different formalisms does also hold also when dealing with bivectors and 3-vectors. More precisely, the following expressions hold:
\begin{equation}\label{SU4SL2H}
  \left\{
     \begin{array}{ll}
         SU(4):\;  B^{A}_{\;\;B} = \chi_{\mf{a}}^{\;A}\,\zeta^{\mf{b}}_{\;B} \quad &\Rightarrow \quad  SL(2;\mathbb{H}):\;
B=  \chi_{\mf{a}}\,\bl{g}\,\zeta^{\mf{b}^\dagger} \;, \\
 SU(4):\; T^{+\,AB} = \chi_{(\mf{a}}^{\;A}\,\chi_{\mf{b})}^{\;B} \quad &\Rightarrow \quad  SL(2;\mathbb{H}):\;
T^{+} = \chi_{(\mf{a}} \, \bl{g}\,\chi_{\mf{b})}^{\dagger} \; \\
         SU(4):\; T^{-}_{AB} = \zeta^{(\mf{a}}_{\;A}\,\zeta^{\mf{b})}_{\;B} \quad &\Rightarrow \quad  SL(2;\mathbb{H}):\;
 T^-  = \zeta^{(\mf{a}}\,\bl{g} \, \zeta^{\mf{b}) \dagger} \;,
    \end{array}
   \right.
\end{equation}
Thus, for instance, as shown in Eq. (\ref{BivectorsSU4}), the bivector $(e_1\wedge e_2)^{\mu\nu}$ is represented in the $SU(4)$ formalism by $\chi_1^{\;A}\zeta^4_{\;B}$. Our claim in Eq. (\ref{SU4SL2H}) is that the $SL(2;\mathbb{H})$ representation of $(e_1\wedge e_2)^{\mu\nu}$ should then be given by
\begin{equation}\label{e1e21}
  e_1\wedge e_2 = \,\chi_{\mf{1}}\,\bl{g}\,\zeta^{\mf{4}^\dagger} = \frac{2}{g_2}\left[
                                                                       \begin{array}{c}
                                                                         \bl{i\,s} \\
                                                                         0 \\
                                                                       \end{array}
                                                                     \right]\,\bl{g}\,
\big[ 0 \;\;\; \bl{is}\big] =
\left[
                      \begin{array}{cc}
                        0 &  \bl{i} - \bl{J} \\
                        0 & 0 \\
                      \end{array}
                    \right] \,.
\end{equation}
On the other hand, as discussed in Sec. \ref{Sec.Bivec3Vec}, in the $SL(2;\mathbb{H})$ approach the bivector $(e_1\wedge e_2)^{\mu\nu}$ should be represented by the following matrix:
\begin{equation}\label{e1e22}
  e_1\wedge e_2 = e_1\,\widetilde{e}_2 - e_2\,\widetilde{e}_1 \,,
\end{equation}
So, using the expressions
$$ e_1 = \frac{1}{2}(\sigma_5+\sigma_0) \;,\;\; \widetilde{e}_1 = \frac{1}{2}(\sigma_5-\sigma_0) \;,\;\; e_2=\widetilde{e}_2 = \frac{1}{2}(\sigma_2+\bl{i}\sigma_4) \,,
 $$
one eventually find that the matrix $e_1\wedge e_2$ is given by
\begin{equation}\label{e1e23}
  e_1\wedge e_2 = \frac{1}{2}(\sigma_5 + \sigma_0)(\sigma_2+\bl{i}\sigma_4) =\left[
                      \begin{array}{cc}
                        0 & \bl{i} - \bl{J} \\
                        0 & 0 \\
                      \end{array}
                    \right]\,,
\end{equation}
which is in perfect accordance with Eq. (\ref{e1e21}). Likewise, using the procedure stated in Eq. (\ref{SU4SL2H}) we can compute the representation of 3-vectors in the $SL(2;\mathbb{H})$ approach. For instance, as shown in Eq. (\ref{3-vectorsSU4}), the self-dual part of the
3-vector $(e_1\wedge e_2 \wedge e_3)^{\mu\nu\lambda}$ is represented in the $SU(4)$ spinorial formalism by $-3\chi_1^{\;A}\chi_1^{\;B}$, whereas its anti-self-dual part vanishes. Therefore, we should have that
\begin{equation}\label{3VEC1}
  \left\{
       \begin{array}{ll}
         (e_1\wedge e_2 \wedge e_3)^+ = -3\chi_1\,\bl{g}\,\chi_1^\dagger =
-3\left[
  \begin{array}{cc}
    \bl{I-i\,K} & 0 \\
    0 & 0 \\
  \end{array}
\right] \,,\\
   (e_1\wedge e_2 \wedge e_3)^- = \left[
  \begin{array}{cc}
    0 & 0 \\
    0 & 0 \\
  \end{array}
\right] \,.
       \end{array}
     \right.
\end{equation}
On the other hand, we have seen in Sec. (\ref{Sec.Bivec3Vec}) that the representation of such a 3-vector in the $SL(2;\mathbb{H})$ formalism should be given by:
\begin{equation*}
  \left\{
     \begin{array}{ll}
       (e_1\wedge e_2 \wedge e_3)^{+} &= e_1 \widetilde{e}_2  e_3  - e_3 \widetilde{e}_2  e_1 + e_2 \widetilde{e}_3  e_1  - e_1 \widetilde{e}_3  e_2 +
e_3 \widetilde{e}_1  e_2  - e_2 \widetilde{e}_1  e_3  \,, \\
\,(e_1\wedge e_2 \wedge e_3)^{-} &=  \widetilde{e}_1 e_2  \widetilde{e}_3  - \widetilde{e}_3 e_2  \widetilde{e}_1 + \widetilde{e}_2 e_3  \widetilde{e}_1  - \widetilde{e}_1 e_3  \widetilde{e}_2 +
\widetilde{e}_3 e_1  \widetilde{e}_2  - \widetilde{e}_2 e_1  \widetilde{e}_3 \,.
\end{array}   \right.
\end{equation*}
Inserting the explicit expressions for the matrices $e_i$ and $\widetilde{e}_i$ in the latter equations we eventually find that
$$ (e_1\wedge e_2 \wedge e_3)^{+}  =   -3\left[
  \begin{array}{cc}
    \bl{I-i\,K} & 0 \\
    0 & 0 \\
  \end{array}
\right] \;\; \text{ and } \;\; (e_1\wedge e_2 \wedge e_3)^{-}  =   \left[
  \begin{array}{cc}
    0 & 0 \\
    0 & 0 \\
  \end{array}
\right] \,, $$
which is in full agreement with Eq. (\ref{3VEC1}).

In the $SU(4)$ approach, the spinors of positive and negative chirality can be contracted yielding a scalar. In particular, the contraction of the frame spinors gives
\begin{equation}\label{ChiZetaContrac}
  \zeta^{\mf{a}}_{\;A} \, \chi_{\mf{b}}^{\;A} = \delta_{\mf{b}}^{\;\mf{a}} \,.
\end{equation}
Therefore, an analogous relation should be valid for our frame. Taking into account the recipe given in Eq. (\ref{SU4SL2H}), there are two natural ways to generate a number from the $SL(2;\mathbb{H})$ frame spinors $\zeta^{\mf{a}}$ and $\chi_{\mf{b}}$, namely
$$ \zeta^{\mf{a}\,\dagger}\,\bl{g}\,  \chi_{\mf{b}}  \quad \text{ and } \quad \chi_{\mf{b}}^\dagger\,\bl{g}\, \zeta^{\mf{a}}  \,.  $$
However, the latter matrix products will, in general, be a quaternion rather that a real or a complex number. Therefore, in order to make contact with Eq. (\ref{ChiZetaContrac}) we should take the real part of such products. For instance, taking the real part of $ \zeta^{\mf{a}\dagger} \bl{g} \,\chi_{\mf{b}} $ yields:
\begin{equation}\label{Contrac1}
  \mf{R}\left[    \zeta^{\mf{a}\,\dagger}\,\bl{g} \,\chi_{\mf{b}}  \right] = \frac{1}{2} \left(  \zeta^{\mf{a}\,\dagger}\,\bl{g} \,\chi_{\mf{b}} +
\chi_{\mf{b}}^\dagger\,\bar{\bl{g}} \,\zeta^{\mf{a}}   \right) =
\frac{1}{2} \left(  \zeta^{\mf{a}\,\dagger}\,\bl{g} \,\chi_{\mf{b}} - \chi_{\mf{b}}^\dagger\,\bl{g} \,\zeta^{\mf{a}}   \right)\,,
\end{equation}
where in the above equation $\mf{R}[\bl{a}]$ stands for the real part of the quaternion $\bl{a}$, i.e. $\mf{R}[\bl{a}]=\frac{1}{2}(\bl{a}+\bar{\bl{a}})$. It has also been used that if $\bl{a}\in\mathbb{H}$ then $\bl{a}^\dagger = \bar{\bl{a}}$ and that $\bar{\bl{g}}=-\bl{g}$. It turns out, however, that the expression (\ref{Contrac1}) is not invariant by the action of the spin group and, therefore, cannot be the $SL(2;\mathbb{H})$ counterpart of  $\zeta^{\mf{a}}_{A}\chi_{\mf{b}}^{A}$. Indeed, applying an infinitesimal $SL(2;\mathbb{H})$ transformation, $Q = \mathbb{I}_2 + \kappa \mathcal{L}$, on the spinors $\zeta^{\mf{a}}$ and $\chi_{\mf{b}}$, it follows that the variation on the right hand side of Eq. (\ref{Contrac1}), up to first order in $\kappa$, is given by
$$ \frac{\kappa}{2} \left[  \zeta^{\mf{a}\,\dagger}\,(\mathcal{L}  \bl{g} -  \bl{g}\mathcal{L} ) \,\chi_{\mf{b}}  +
\chi_{\mf{b}}^\dagger\,(\mathcal{L}^\dagger \bl{g} -  \bl{g}\mathcal{L}^\dagger) \,\zeta^{\mf{a}} \right]  \,,$$
which does not vanish for a general element of the Lie algebra $\mathcal{L}$. Rather, in order to generate the desired result, we should use the contractions
$$ \zeta^{\mf{a}\,\dagger}\,  \chi_{\mf{b}}\,\bl{g} \quad \text{ and } \quad \chi_{\mf{b}}^\dagger\,  \zeta^{\mf{a}}\,\bl{g}  \,,  $$
with $\bl{g}$ at the right hand side instead of being in the middle of the spinors. Indeed, the latter contractions are easily seen to be invariant under the action of the spin group, i.e. they are true scalars. Moreover, computing the real part of $\zeta^{\mf{a}\,\dagger}  \chi_{\mf{b}} \bl{g}$ we have
$$  \mf{R}\left[  \zeta^{\mf{a}\,\dagger}\,  \chi_{\mf{b}}\,\bl{g} \right]  = \delta_{\mf{b}}^{\;\mf{a}} \,, $$
which is the desired analogous relation (\ref{Contrac1}). With this we have completed the bridge between the two spinorial formalisms for six-dimensional Lorentzian spaces.


\section{Conclusions}

In this work we have shown how vectors, bivectors and 3-vectors of six-dimensional spacetimes are represented in the $SL(2;\mathbb{H})$ spinorial formalism. Vectors and bivectors had already been considered in this formalism in Ref. \cite{Kugo}, but here we follow step by step approach in which the proper tensor representations carried by these geometrical objects is derived, rather than stated. Moreover, we have pointed out that general tensorial objects built from the fundamental representations of a quaternionic group do not carry a representation of the group, contrary to what happens when groups are represented in vector spaces over real and complex fields. In spite of the latter observation being quite basic and of broad application for quaternionic groups, the authors are not aware of any previous acknowledgement of this fact in the literature. In particular, this implies that simple tensorial objects in  six-dimensional spacetimes like a rank two symmetric and trace-less tensor and a rank four tensor with the symmetries of the Weyl tensor, both of which carry irreducible representations of the Lorentz group $SO(5,1)$, cannot be represented in this quaternionic spinorial formalism (at least not in a covariant form, without introducing a frame of spinors). Therefore, we conclude that the two-component spinorial formalism in six dimensions is not as advantageous as the two-component spinorial formalism in four-dimensional spacetimes. Rather, we believe that for most applications the four-component spinorial formalism, based on the $SU(4)$ spin group, is generally much more handy and could lead to more geometrical insights than the $SL(2;\mathbb{H})$ spinorial formalism presented here. The difficulties on the use of the quaternionic spinorial formalism can be traced back to the fact that when we write a spinor as a two-component object it follows that each component is a quaternion, and each quaternion is associated to a pair of complex numbers. Thus, it is like each component of the spinor field is actually a field itself with two complex degrees of freedom. This is why the index formalism, as the one developed by R. Penrose and W. Rindler in four dimensions \cite{Penrose84}, does not work in six dimensions.

In spite of what was argued in the previous paragraph, there might exist some specific applications in which the quaternionic method can be fruitful and even more elegant than the $SU(4)$ approach. This is why it is important to known how to connect the $SL(2;\mathbb{H})$ and the $SU(4)$ approaches, which was done here in a covariant way for the first time in the literature. A more rudimentary way of connecting these two spinorial approaches was also pointed out in Ref. \cite{Kugo}, but there a specific matrix representation of the quaternions is used for this purpose. In particular, here we have seen the relevance of the quaternion $\bl{g}$ (which obeys $\bar{\bl{g}}=-\bl{g}$ and $\bl{g}^2=0$) in the transition between the two spinorial approaches, whereas in Ref. \cite{Kugo} this object does appear.
An example in which the $SL(2;\mathbb{H})$ approach might be more natural, and therefore more useful, than the $SU(4)$ approach is in the study of the
the conformal group in four-dimensional spacetimes \cite{Wilker}. The reasoning is that the Lorentz group is four dimensions, $SO(3,1)$, is contained in $SL(2;\mathbb{C})$, while the conformal group in four dimensions is given by $SO(4,2)$, which by means of complexification can be associated to $SO(5,1)$ which is contained $SL(2;\mathbb{H})$. Since $SL(2;\mathbb{C})$ is trivially contained in $SL(2;\mathbb{H})$, it follows that through the quaternionic approach we can tackle the conformal group in four dimensions with a clear distinction between the Lorentz group and the special conformal transformations. Of course we could also consider the conformal group in terms of $SU(4)$ spinors, but in this formalism the subgroup of Lorentz transformations is not easily and naturally identified. We intend to discuss about the four-dimensional conformal group using the $SL(2;\mathbb{H})$ formalism in a forthcoming work. Note that the complexification of the Minkowski space performed here is of central importance in order to put forward the latter research project. As far as we know, such complexification has not been discussed elsewhere.

Another advance accomplished by the present article is the physical interpretation of the $SL(2;\mathbb{H})$ transformations, which has been achieved here by means of studying the Lie algebra of this spin group. It is worth stressing that the Lie algebra of $SL(2;\mathbb{H})$ has also been considered in Ref. \cite{Lukierski}, see also \cite{Rocha} for another signature, but the interpretation of the generators of the group have not been given previously.

In a forthcoming paper we shall follow similar steps in ten dimensions, using the fact that in this case the chiral spinors can be represented by two-component objects with entries on octonions \cite{Kugo,Baez:2010bbj}.

\section*{Acknowledgments}

C. B. would like to thank Conselho Nacional de Desenvolvimento Cient\'{\i}fico e Tecnol\'ogico (CNPq) for the partial financial support through the research productivity fellowship. Likewise,  C. B. thanks Universidade Federal de Pernambuco for the funding through Qualis A project.  J. V. thanks CNPq for the financial support. We both thank CAPES for the invaluable funding of the graduation program of our department. We are grateful to Urs Schreiber for pointing out the noteworthy reference \cite{Baez:2010bbj} after we released the preprint. In addition, we thank the kind and enlightening e-mail of Paul Townsend mentioning Dirac's hidden article \cite{Dirac}.



\end{document}